\documentclass{article}
\usepackage{a4wide}
\usepackage{amsmath}
\usepackage{amscd}
\usepackage{amsfonts}
\usepackage{amssymb}
\usepackage{amsthm}
\usepackage{epsfig}
\usepackage{graphicx}
\newcommand{\ro}{\mathrm}
\newcommand{\C}{\mathcal{C}}
\newcommand{\real}{\mathbb{R}}
\newcommand{\PP}{\mathbb{P}}

\newcommand{\ps}[1][t]{\Psi_{#1}}
\newcommand{\psout}[1][t]{\Psi_{#1}^{\rm out}}
\newcommand{\psac}[1][t]{\Psi_{#1}^{\rm ac}}
\newcommand{\pspp}[1][t]{\Psi_{#1}^{\rm \scriptscriptstyle pp}}
\newcommand{\psat}[1][out]{\hat{\Psi}_0^{\rm #1}}
\newcommand{\xt}[1][t]{\frac{q}{#1}}
\newcommand{\tx}[1][q]{\frac{t}{|q|}}
\newcommand{\nhalf}[1][3]{\frac{#1}{2}}
\newcommand{\Qo}{q_0}
\newcommand{\Q}[1][t]{\ro{Q}(\Qo,#1)}
\newcommand{\Qp}[1][t]{\ro{Q}(q,#1)}
\newcommand{\qt}[1][t]{\frac{\ro{Q}(\Qo,#1)}{#1}}
\newcommand{\qtp}[1][t]{\frac{\ro{Q}(q,#1)}{#1}}
\newcommand{\vpsi}[1][]{v^{\Psi^{#1}}}
\newcommand{\setonea}[1][B]{#1_{\delta_1ab}}
\newcommand{\settwoa}[1][B]{#1_{\delta_1 \delta_2ab}}
\newcommand{\eps}{\varepsilon}
\newcommand{\U}[1][\delta_2]{U_{#1}}
\newcommand{\A}[1][\gamma]{\ro{A}_{#1}}

\newcommand{\M}{\ro{M}}
\newcommand{\ph}[1][+]{\Phi_{#1}}
\newcommand{\R}{R_{a,T}(t)}

\newcommand{\Pgamma}[1][]{\ro{P}^{\rm #1}_\gamma}
\renewcommand{\Im}{\ro{Im}}
\renewcommand{\Re}{\ro{Re}}
\renewcommand{\phi}{\varphi}
\newcommand{\deltat}{\Delta T}
\newcommand{\D}[1][\alpha]{\mathcal{D}_{#1}}
\renewcommand{\hat}{\widehat}
\renewcommand{\tilde}{\widetilde}
\newcommand{\G}{\ro{G}}
\theoremstyle{plain}
\newtheorem{thm}{Theorem}
\newtheorem{lem}{Lemma}
\newtheorem{cor}{Corollary}
\newtheorem{defi}{Definition}

\theoremstyle{definition}
\newtheorem{rem}{Remark}

\newenvironment{pro}[1][]{\noindent{\bf Proof #1.}}{\hfill $\Box$}

\begin{document}
\begin{center}
{\Large\bf Asymptotic Behavior of Bohmian Trajectories in\\Scattering Situations}\\
\end{center}
\vspace{1cm}
S. R\"omer, D. D\"urr, T. Moser\\
\\
\noindent Mathematisches Institut der Ludwig-Maximilians-Universit\"at M\"unchen,\\
Theresienstra\ss e 39, 80333 M\"unchen, Germany\\
e-mail: roemer@mathematik.uni-muenchen.de
\vspace{1cm}
\begin{small}
\begin{center}{\bf Abstract.}
\end{center}
We study the asymptotic behavior of Bohmian trajectories in a scattering situation with short range potential $V$ and for
wave functions $\ps[]=\psac[]+\pspp[]\in L^2(\real^3)$ with a scattering and a bound part. It is shown that the set of
possible
trajectories splits into trajectories whose long time behavior is governed by the scattering part $\psac[]$
of the wave function (scattering trajectories)
and trajectories whose long time behavior is governed by the bound part $\pspp[]$ of the wave function
(bound trajectories).
Furthermore, the scattering trajectories behave like trajectories in
classical mechanics in the limit $t\rightarrow\infty$. As an intermediate step we show that the asymptotic velocity
$v_\infty:=\lim\limits_{t\rightarrow\infty}Q/t$ exists almost surely and is randomly distributed with the
density $|\hat{\ps[]}^{\rm out}|^2$, where $\psout[]$ is the outgoing asymptote of the scattering part of the wave function.
\end{small}

\section{Introduction}
\label{sec.intro}

Bohmian mechanics \cite{bohm:52, bell:87, duerr:01, duerr3:04, duerr:92}
is a theory of particles in motion that is experimentally equivalent to quantum mechanics
whenever the latter makes unambiguous predictions \cite{duerr3:04}.
While Bohmian trajectories are in general highly non-Newtonian, we will show that, in the special
context of potential scattering theory, the long time asymptotes of the trajectories associated
with scattering wave functions are classical straight lines with an asymptotic velocity
$v_\infty$ that is randomly distributed with the density $|\hat{\ps[]}^{\ro{out}}|^2$, where $\psout[]$ is the
outgoing asymptote of the wave function and $\;\hat\;\;$ denotes Fourier transformation.\\

\noindent In Bohmian mechanics, the state of a spinless, non-relativistic particle is described by its (normalized) quantum
mechanical wave function $\ps(q)$, where $q\in\real^3$, and by its actual configuration (its position)
$Q\in\real^3$.\\
The wave function evolves according to the Schr\"{o}dinger equation
\begin{equation}
\label{eq.schroedinger}
i\hbar\frac{\partial\ps}{\partial t}=H\ps
\end{equation}
and governs the motion of the particle by
\begin{equation}
\label{eq.eq-of-motion}
\frac{d\,Q}{dt}=\vpsi(Q,t):=\frac{\hbar}{m}\Im\left(\frac{\nabla\ps(Q)}{\ps(Q)}\right).
\end{equation}
Here $m$ is the mass of the particle. In \eqref{eq.schroedinger}
$H$ is the usual non-relativistic Schr\"{o}dinger Hamiltonian
\begin{equation}
\label{eq.hamiltonian}
H=-\frac{1}{2m}\triangle+V(q)=:H_0+V(q)
\end{equation}
with the non-relativistic interaction potential V\footnote{More
rigorously: $H$ is a self-adjoint extension of $H|_{C
^\infty_{0}(\Omega)}=-\frac{1}{2m}\triangle+V$ (with $V:\Omega\subset\real^3\rightarrow\real)$ on the Hilbert space $L^2(\real^3)$
with domain $\mathcal{D}(H)$. See Definition \ref{def.Vn}.}. From now on we shall use natural units $m=\hbar=1$.\\
For a wave function $\ps[]$ the actual configuration $\ro{Q}$ is randomly distributed according to the equivariant
probability measure $\PP^{\ps[]}$ on configuration space given by the density $|\ps[](q)|^2$ (Born's statistical
rule); see \cite{duerr:92}. Roughly speaking this means that a typical Bohmian trajectory will always stay in the
main part of the support of $\ps[]$ (see Subsection \ref{subsec.global-existence}).\\

\noindent We shall look at scattering situations where $V$ is a sufficiently smooth short range potential falling off
like $|q|^{-4-\eps}$ for some $\eps>0$ and $|q|\rightarrow\infty$. This of course includes the case
of free motion ($V\equiv0$).\\
For scattering wave functions $\psac$ in $\mathcal{H}_{\rm ac}(H)$, the absolute continuous spectral subspace,
we show that $\PP^{\psac[0]}$-almost all Bohmian trajectories behave like classical
(Newtonian) trajectories for $t\rightarrow\infty$, i.e. their long time asymptotes are
straight lines with a uniform velocity $v_\infty=\lim\limits_{t\rightarrow\infty}\frac{Q(t)}{t}$.
In accordance with orthodox quantum mechanics, we find that $v_\infty$ is randomly distributed with density
$|\psat(\cdot)|^2$, where $\psout$ is the outgoing asymptote of $\psac$.
We shall use the terminology "straight line motion" for motion with
uniform velocity.\\
We give a heuristic argument why this should be so:
It is known (see Lemma \ref{lem.est-psac-phi1}) that the long time limit (in $L^2$-sense)
of a scattering wave function $\psac$ is a local plane wave $\phi_1=(it)^{-\nhalf}\exp\left(i\frac{q^2}{2t}\right)
\psat\left(\xt\right)$. So the support of $\psac$
essentially moves out to spatial infinity linear in time. But then a typical Bohmian trajectory $\ro{Q}$
(that stays in the main part of the support of $\psac$) will
move out to infinity linear in time, too, that is $\frac{\ro{Q}}{t}=\mathcal{O}(1)$ for large times $t$. An
estimate on the asymptotic behavior of the velocity $\vpsi[\ro{ac}]\left(\ro{Q},t\right)$ of such a typical
trajectory $\ro{Q}$ is provided by
\begin{equation}
\label{eq.vpsac-approx-vphi1}
v^{\phi_1}\left(\ro{Q},t\right)=
        \Im\left(\frac{\nabla\phi_1\left(\ro{Q},t\right)}{\phi_1\left(\ro{Q},t\right)}\right)
\end{equation}
Rewriting $\phi_1$ in complex polar coordinates, i.e.
\begin{equation*}
\phi_1(q,t)=\ro{R}(q,t)e^{i\left(\frac{q^2}{2t}+\ro{S}(\xt)\right)}
\end{equation*}
with R and S real valued, and keeping in mind that $\frac{\ro{Q}}{t}=\mathcal{O}(1)$ we get
\begin{equation}
\label{eq.vpsac-approx-qt}
\vpsi[\ro{ac}]\left(\ro{Q},t\right)\stackrel{t\to\infty}{=}v^{\phi_1}\left(\ro{Q},t\right)=
    \frac{\ro{Q}}{t}+\frac{1}{t}\nabla_kS(k)\mid_{k=\frac{\ro{Q}}{t}}\stackrel{t\to\infty}{=}\frac{\ro{Q}}{t}\,.
\end{equation}
But $\frac{d\ro{Q}}{dt}=v(\ro{Q},t)=\frac{\ro{Q}}{t}$ defines straight line motion.\\

\begin{figure}[h]
\begin{center}
\epsfxsize=6cm
\epsffile{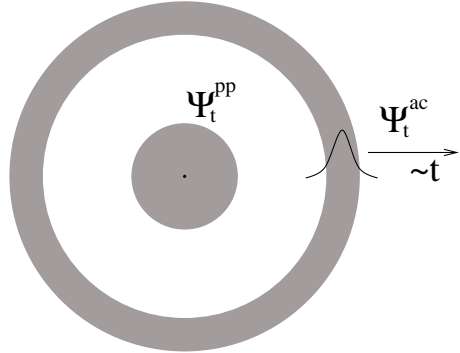}
\end{center}
\caption{\footnotesize Splitting of the support of $\ps=\psac+\pspp$ for large times $t$.}
\label{fig.supportwf}
\end{figure}

\begin{figure}[ht]
\begin{center}
\epsfxsize=10cm
\epsffile{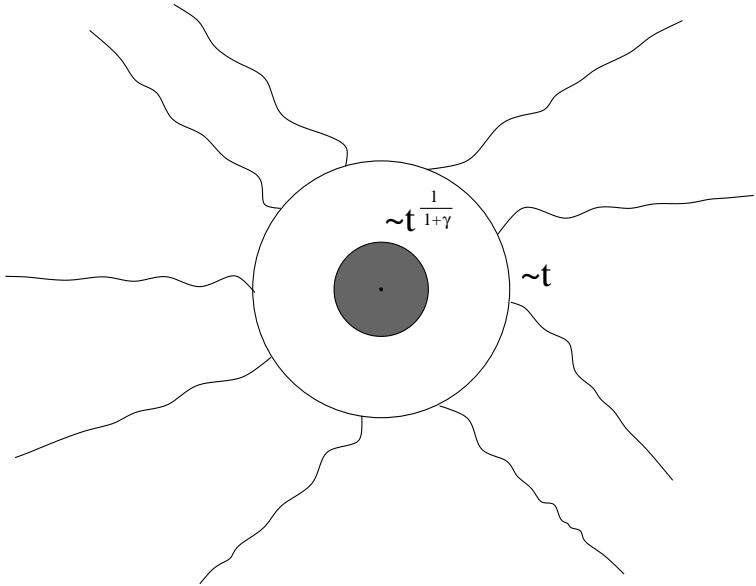}
\end{center}
\caption{\footnotesize Splitting of the Bohmian trajectories made by $\ps=\psac+\pspp$  for large times $t$.
Scattering trajectories stay outside some ball with radius growing linear in time ($\sim t$) and become straight
lines asymptotically. Bound trajectories stay inside some ball growing only sublinear in time ($\sim
t^{\frac{1}{1+\gamma}}$ for some suitable $\gamma>0$).}
\label{fig.pp-split}
\end{figure}

\noindent Moreover, we show that classical behavior of Bohmian trajectories in the limit $t\rightarrow\infty$
arises also for wave functions $\ps=\pspp+\psac$ with a bound part $\pspp$ in $\mathcal{H}_{\rm pp}(H)$,
the pure point spectral subspace: We prove that $\PP^{\ps[0]}$-almost all Bohmian trajectories are either
(with probability $\|\psac[0]\|^2$) trajectories whose long time behavior is governed by the scattering part
$\psac$ of the wave function (\emph{scattering} trajectories)
or trajectories (with probability $\|\pspp[0]\|^2$) whose long time behavior is governed by the bound part
$\pspp$ (\emph{bound} trajectories). Since the Bohmian equation of motion
\eqref{eq.eq-of-motion} is not linear in $\ps[]$ this is not a trivial result. However, it is clear heuristically.\\
On the one hand, it is known (see e.g.
\cite{ruelle:69, perry:83}) that the spatial support of the bound part $\pspp$ of the wave function stays
concentrated around the origin (the scattering center) for all times t. Since on the other hand, the support of
the scattering part $\psac$ of the wave function essentially moves out to infinity linear in time, at large
times t there will be two distinct parts of the support of the whole wave function. In Figure \ref{fig.supportwf},
we drew the situation for the case that the support of the outgoing asymptote in momentum space $\psat$ is mainly
concentrated away from zero.
Note, however, that what we said above is true even for the case where $\psat$ is mainly concentrated around
zero, since the only part of the support of $\psac$ that stays close to the scattering center for all times is that
corresponding to $\psat(k)$ where $k$ is \emph{exactly} zero.\\
So a typical Bohmian trajectory should either stay bound or again move out to
infinity linear in time. Moreover, in the first case the scattering part of the wave function and
in the second case the bound part of the wave function will be negligible.
This already gives us the splitting into bound and scattering trajectories
and consequently the asymptotic classical behavior of the scattering trajectories (see Figure \ref{fig.pp-split}).\\

\noindent Note that, since a bound wave function $\pspp$ stays in
the sphere of influence of the potential $V$ even in the long time limit, it should depend on the exact form of the
potential $V$ and on $\pspp$ itself whether bound trajectories behave like classical
trajectories or not. This is a question that we will not deal with here.\\
We show, however, that bound trajectories stay inside some
ball around the origin with radius growing sublinear in time, that is we prove that they move
out to spatial infinity on a much larger time scale than scattering trajectories. While this suffices for getting
the afore mentioned splitting of trajectories and thus the classical behavior of scattering trajectories it is surely not
the best possible result on bound trajectories one can expect. We shall deal with the
behavior of bound trajectories in more detail in a subsequent work.\\
Another open question is how one could characterize the sets of initial configurations that lead to bound resp.
scattering trajectories. Are they open or closed or neither of both? Are the starting points of bound trajectories
intermixed with those of scattering ones or do they form well discernible sets? How does this depend on the
dimension (or the symmetry) of the problem?\\
Finally, what about more general (scattering or non-scattering) situations? When do Bohmian trajectories look like classical ones
\emph {in general}? We consider this question to be the key question of the classical limit in Bohmian mechanics
\cite{allori:02}. It is our conviction  that the methods developed in this article are naturally fit to give
mathematically rigorous results also in the general case and plan to use them to just that end in the future.\\

\noindent The problem of establishing what intuitively seems clear, that asymptotically particles move freely on
straight lines (for short range potentials), has been addressed before by Shucker \cite{shucker:80}, Biler \cite{biler:84} and Carlen \cite{carlen:84, carlen:85} for stochastic mechanics. Although Shucker proved results for $V\equiv0$ only and from those results one cannot infer the existence of
an asymptotic velocity, steps in his proof are also useful for our case\footnote{A paraphrase of his results for Bohmian Mechanics and an appraisal from the viewpoint of Bohmian Mechanics can be found in \cite{daumer:94} p. 48.}. Biler used the methods of Shucker to treat potential scattering in one dimension for scattering wave functions with "momentum" supported compactly in $(0,\infty)$. The general $3$-dimensional case (for pure scattering wave functions) was treated by Carlen, who used methods relying on $L_2$-estimates rather than the pointwise estimates of Shucker. However, contrary to those of Shucker, his methods can't be extended to give the convergence of the real velocity $\vpsi$ to $v_\infty$.\\

\noindent The article is organized as follows. First we set up the mathematical framework (Section \ref{sec.math-frame}).
We give a brief account of equivariance (Subsection
\ref{subsec.global-existence}) and list some results of potential scattering theory (Subsection \ref{subsec.pot-scat}).
In Section \ref{sec.as-behav} we state our results on the asymptotic behavior of Bohmian trajectories for pure scattering
wave functions and for general wave functions $\ps=\psac+\pspp$
(Theorem \ref{thm.pure-scattering} and Corollary \ref{cor.straight-pure} resp.
Theorem \ref{thm.mixed} and Corollary \ref{cor.straight-mixed}). Section \ref{sec.proof} contains the proof.

%-----------------------------------------------------------------------------------

\section{Mathematical Framework}
\label{sec.math-frame}

\subsection{Equivariance}
\label{subsec.global-existence}

The dynamical system defined by Bohmian mechanics is naturally associated with a family of finite measures $\PP^{\ps}$
given by the densities
$\rho^{\ps}(q):=|\ps(q)|^2$ on configuration space $\real^3$.
Let $\ph[t,t_0]:\real^3\rightarrow\real^3$ be the flow map of \eqref{eq.eq-of-motion}, i.e., if $q$ is
the initial configuration at time $t_0$, $\ph[t,t_0](q)$ is the configuration at time $t$ which $q$ is transported to by
\eqref{eq.eq-of-motion}. Then the density $\rho_{t_0}:=\rho^{\ps[t_0]}$ is transported to
$\rho_t=\mathcal{F}_{t,t_0}\left(\rho_{t_0}\right):=\rho_{t_0}\circ\ph[t,t_0]^{-1}$.
We say that the functional $\ps\mapsto\PP^{\ps}$, from wave functions to the finite measures
$\PP^{\ps}$ (given by the densities $\rho^{\ps})$ on configuration space, is equivariant if the diagram
\begin{equation*}
\begin{CD}
\ps[t_0]@>U_{t-t_0}>>\ps\\
@VVV @VVV\\
\rho^{\ps[t_0]}@>>\mathcal{F}_{t,t_0}>\rho^{\ps}
\end{CD}
\end{equation*}
commutes \cite{duerr:92}, i.e. $\rho_{t}=\rho^{\ps[t]}$ for all times $t$. Here $U_t=e^{-iHt}$ is the solution map
for the Schr\"{o}dinger equation
\eqref{eq.schroedinger} and $\mathcal{F}_{t,t_0}$ is the solution map for the natural evolution on densities
arising from \eqref{eq.eq-of-motion} (see above).\\
On the family of measures $\PP^{\ps}$ we bestow the role usually played
by the stationary "equilibrium measure" \cite{duerr:92}. Thus $\PP
^{\ps}$ defines our notion of typicality, which by equivariance is time independent:
\begin{equation}
\label{eq.equivariance}
\begin{split}
\PP^{\ps[t_1]}(A)&=\int\limits_{\real^n}\chi_A(q)|\ps[t_1](q)|^2\,dq=
\int\limits_{\real^n}\big(\chi_{\ph[t_2,t_1](A)}\cdot\ph[t_1,t_2]\big)(q)|\ps[t_1](q)|^2\,dq=\\
&=\int\limits_{\real^n}\chi_{\ph[t_2,t_1](A)}(q)\big(|\ps[t_1]|^2\cdot\ph[t_2,t_1]\big)(q)\,dq=\\
&=\int\limits_{\real^n}\chi_{\ph[t_2,t_1](A)}(q)|\ps[t_2](q)|^2\,dq=\PP^{\ps[t_2]}\big(\ph[t_2,t_1](A)\big),
\end{split}
\end{equation}
for all measurable sets $A\subset\real^3$. Here $\chi_A$ denotes the characteristic function of $A$.\\
From now on we will write
\begin{equation*}
\Q:=\ph[t,0](\Qo)\qquad(t\in\real\,,\;\Qo\in\real^3)
\end{equation*}
for the solution of \eqref{eq.eq-of-motion} with initial configuration $\Qo$\footnote{
Without loss of generality we have set $t_0=0$.}.
%-------------------------------------------------------------------------------------------------------------------

\subsection{Potential Scattering Theory}
\label{subsec.pot-scat}

We look at a scattering situation described by a Hamiltonian $H=H_0+V,\;\mathcal{D}(H)\subset
L^2(\real^3)$, that is by a self adjoint extension on $L^2(\real^3)$ of $\tilde H=-\nhalf[1]\triangle+V,
\;\mathcal{D}(\tilde H)= \rm C_0^\infty(\Omega)$, where $\Omega=\{q\in\real^3\mid V(q) \mbox{ is not singular}\}$
and $V$ is a short-range potential, $V\in(V)_n$ ($n\geq2$):

\begin{defi}
\label{def.Vn} For $n\geq 2$ the following conditions on the
potential $V$ will be denoted by $V\in (V)_n$.
\begin{enumerate}
\item $V \in L^2(\real^3,\real)\,.$
\item $V$ is $\ro C^\infty$ except, perhaps, at finitely many singularities.
\item There exist $\eps_0>0$, $C_0>0$ and $R_0>0$ such that $|V(q)|\leq C_0\langle q\rangle^{-n-\eps_0}$ for
all $|q|\geq R_0$.\\
Here $\langle q\rangle:=\left(1+q^2\right)^{\nhalf[1]}$.
\end{enumerate}
\end{defi}

\noindent Clearly the wave operators $W_{\pm}:=\underset{t\rightarrow \pm
\infty}{\ro{s}-\lim}e^{iHt}e^{-iH_0t}$ exist\footnote{Here $\ro{s}-\lim$ denotes the limit in $L^2$-sense.}
and are asymptotically
complete (see e.g. \cite{ikebe:60}). $W_\pm$ are called asymptotically complete if their range fulfills
\begin{equation*}
\ro{Ran}W_\pm=\mathcal{H}_{\rm c}(H)=\mathcal{H}_{\rm ac}(H)\,,
\end{equation*}
where $\mathcal{H}_{\rm c}(H)$ resp. $\mathcal{H}_{\rm ac}(H)$ denotes the spectral subspace of $L^2(\real^3)$ that belongs
to the continuous resp. the absolutely continuous spectrum of the hamiltonian $H$. Since $L^2(\real^3)$ is the
orthogonal sum of $\mathcal{H}_{\rm c}(H)$ and $\mathcal{H}_{\rm pp}(H)$ (the subspace that belongs to the pure
point spectrum of $H$) this implies that a general solution $\ps=e^{-iHt}\ps[0]$ of the Schr\"odinger
equation \eqref{eq.schroedinger} is at all times $t$ given by the unique decomposition $\ps=\psac+\pspp$
into a scattering wave function $\psac\in\mathcal{H}_{\rm ac}(H)$ and a bound wave function $\pspp\in
\mathcal{H}_{\rm pp}(H)$. In addition, all the spectral subspaces are invariant under the full time evolution $e^{-iHt}$,
so $\psac=e^{-iHt}\psac[0]$ and $\pspp=e^{-iHt}\pspp[0]$.\\
More importantly, asymptotic completeness of $W_\pm$ guarantees the existence of a unique outgoing/incomming
asymptote $\ps^{\rm out/in}:=W_\pm^{-1}\psac$ for every scattering wave function $\psac$. Due to
the so called intertwining property, $HW_\pm=W_\pm H_0\,,\;\ps^{\rm out/in}$ evolves
according to the free time evolution $e^{-iH_0t}$.\\

%-------------------------------------------------------------------------------------------------------------------
\section{Asymptotic Behavior of Bohmian Trajectories in\\ Scattering Situations}
\label{sec.as-behav}
We consider Hamiltonians $H=H_0+V,\;\mathcal{D}(H)\subset L^2(\real^3)$, with $V\in(V)_4$.
Moreover, we restrict ourselves to initial wave functions that are $\ro C^\infty$-vectors\footnote{
Some special $\ro C^\infty$-vectors are eigenfunctions and "wave packets" $\Psi\in\ro{Ran}(P_{[E_1,\,E_2]})$
with $P_{[E_1,\,E_2]}$ the spectral projection of $H$ to the finite energy interval $[E_1,\,E_2]$.} of $H,\;
\ps[0]\in\ro C^\infty(H)=\cap_{n=1}^{\infty}\mathcal{D}(H^n)$. Note that $\ro C^\infty(H)$ is a core,
that is a domain of essential self-adjointness of H.\\

\noindent First we shall look at pure scattering wave functions $\ps[0]=\psac[0]\in\mathcal{H}_{\rm ac}(H)$.
We define a convenient subset of $\mathcal{H}_{\rm ac}(H)$ for which we establish our results.\\

\begin{defi}
\label{def.C}
$f:\real^3\rightarrow\mathbb{C}$ is in $\C$ if
\begin{gather*}
f\in\mathcal{H}_{\rm ac}(H)\,\cap\,\ro{C}^\infty(H)\,,\\
%H^nf\in\mathcal{D}(H),\; n\in\{0,1,2\},\\
\langle q\rangle^2H^nf\in L^2(\real^3)\,,\; n\in\{0,1,\ldots, 3\}\,,\\
\langle q\rangle^4H^nf\in L^2(\real^3)\,,\; n\in\{0,1,\ldots, 3\}\,,
\end{gather*}
where again $\langle q\rangle:=\left(1+q^2\right)^{\nhalf[1]}$.
\end{defi}
\hfill\\

\begin{rem}
\label{rem.ex-ps-in-C}
Let $\ro{H}^{m,s}$ be the weighted Sobolev space
\begin{equation*}
\ro{H}^{m,s}:=\left\{f\in L^2(\real^3)\mid\langle q\rangle^s\left(1-\triangle\right)^{\nhalf[m]}
        f\in L^2(\real^3)\right\}\,.
\end{equation*}
Example conditions for which $\ps[0]\in\C$ are
\begin{enumerate}
\item
$V\in(V)_n$ for $n\geq2$ and $\ps[0]\in\mathcal{H}_{\rm ac}(H)\,\cap\,
\ro{C}_0^\infty(\real^3\backslash\mathcal{E})$ where $\mathcal{E}$ denotes the set of singularities of $V$,
\item
$V\in(V)_n$ for $n\geq2$, $V\in\ro{H}^{4,4}$ and
$\ps[0]\in\left(\bigcup\limits_{E>0}\ro{Ran}\left(P_{[0,\,E]}\right)\right)\,\bigcap\,\ro{H}^{6,4}$.
\end{enumerate}
Clearly both sets for $\ps[0]$ are dense in $\mathcal{H}_{\rm ac}(H)$.
\end{rem}
\vspace{0.3cm}

\noindent For $\ps[0]\in\C$ we show the following.

\begin{thm}
\label{thm.pure-scattering}
Let $H=H_0+V$ with $V\in(V)_4$ and let zero be neither an eigenvalue nor a resonance\footnote{
Zero is a resonance of $H$ if there exists a solution $f$ of $Hf=0$ such that
$\langle \cdot\rangle^{-\gamma}f\in L^2(\real^3)$ for any $\gamma>\nhalf[1]\,$ but not for $\gamma=0$
(see e.g. \cite{yajima:95} p.552).
%; see \cite{jensen:79} p.584 for a slightly different definition).
The occurrence of a zero eigenvalue or resonance is an exceptional event: For Hamiltonians
$H(c)=H_0+cV$ the set of parameters $c\in\real$, for which zero is an eigenvalue or a resonance, is discrete
(see e.g. \cite{jensen:79} p.589).
}
of $H$. Let
$\ps[0]\in\C$. Then
\begin{enumerate}
\item
the Bohmian trajectories $\Q$ exist globally in time for $\PP^{\ps[0]}$-almost all initial configurations $\Qo\in\real^3$.
\item
$v_\infty(\Qo):=\lim\limits_{t\rightarrow\infty}\qt$ exists for $\PP^{\ps[0]}$-almost all $\Qo\in\real^3$ and it is
randomly distributed with the density
$|\psat(\cdot)|^2$, i.e. for every measurable set $\ro{A}\subset\real^3$
\begin{equation}
\label{eq.v-infty-distr-pure}
\PP^{\ps[0]}\left(v_\infty\in\ro{A}\right)=
\int\limits_{\ro{A}}|\psat(k)|^2\,d^3k\,.
\end{equation}
\item
For $\PP^{\ps[0]}$-almost all Bohmian trajectories the asymptotic velocity is given by $v_{\infty}$,
i.e. for all $\eps>0$ there exists some $T>0$ and some $C<\infty$ such that
\begin{equation}
\label{eq.P(v-well-behaved)-pure}
\PP^{\ps[0]}\left(\left\{\Qo\in\real^3\mid
    \left|\vpsi\left(\Q,t\right)-v_{\infty}(\Qo)\right|<Ct^{-\nhalf[1]}\quad\forall\,t\geq T\right\}\right)>1-\eps\,.
\end{equation}
\end{enumerate}
\end{thm}

\noindent In the proof we shall use ideas of Shucker \cite{shucker:80}, who, for $V\equiv0$, proved results
equivalent to part (ii) for Nelson's stochastic mechanics.

\begin{rem}
\label{rem.why-V4}
The condition $V\in(V)_4$ is technically related to the expansion in generalized eigenfunctions in \cite{teufel2:99,duerr1:04}.
\end{rem}

\noindent Up to now our results are formulated in terms of the
velocity: The asymptotic velocities of $\PP^{\ps[0]}$-almost all trajectories are those of straight paths.
As an easy corollary to Theorem \ref{thm.pure-scattering} we obtain a statement about the trajectories themselves:
$\PP^{\ps[0]}$-almost every trajectory $\Q$ becomes straight in the sense that from some large time on
it stays close to some straight path for arbitrary long time.

\begin{cor}
\label{cor.straight-pure}
Let $H=H_0+V$ with $V\in(V)_4$ and let zero be neither an eigenvalue nor a resonance of $H$. Let
$\ps[0]\in\C$. Then $\PP^{\ps[0]}$-almost all Bohmian trajectories become straight
lines asymptotically, i. e.
for all $\eps>0$, $\delta>0$ and $\deltat>0$ there exists some $T>0$ such that
\begin{equation}
\label{eq.P(psac-straight)}
\PP^{\ps[0]}\bigg(\bigg\{\Qo\in\real^3\mid \sup\limits_{T'\geq T}\;\sup\limits_{t\in[T',T'+\deltat]}
        \left|\Q-g(\Qo,T',t)\right|<\delta\bigg\}\bigg)>1-\eps.
\end{equation}
Here $g(\Qo,T',t):=\Q[T']+v_\infty(\Qo)(t-T')$ is the straight path a particle with the uniform velocity
$v_\infty(\Qo):=\lim\limits_{t\rightarrow\infty}\qt$ would follow.
\end{cor}

\begin{rem}
\label{rem.why-not-straight-forall-t}
One might be inclined to prove something stronger, namely that
$\PP^{\ps[0]}$-almost every trajectory $\Q$ becomes straight in the sense that from some large time on
it stays close to some straight path \emph{for all} time:\\

\noindent\emph{For $\PP^{\ps[0]}$-almost all $\Qo\in\real^3$ there exists some straight path $g_\infty(\Qo,t)=
g_\infty(\Qo,0)+v_\infty(\Qo)t$ such that}
\begin{equation}
\label{eq.straight-deltaT=infty}
\lim\limits_{t\rightarrow\infty}\left|\Q-g_\infty(\Qo,t)\right|=0\,.
\end{equation}

\noindent However, to get this stronger statement the error in \eqref{eq.P(v-well-behaved)-pure}
would have to fall of faster than $t^{-1}$. But this cannot be achieved generically.
In the introduction (Equation \eqref{eq.vpsac-approx-vphi1} and \eqref{eq.vpsac-approx-qt}), we already saw that
even the velocity field made by the long time asymptote $\phi_1$ of the pure scattering wave function
$\ps=\psac$ is given by
\begin{equation*}
v^{\phi_1}(\ro{Q},t)=\frac{\ro{Q}}{t}+\frac{1}{t}\nabla_kS(k)\big|_{k=\frac{\ro{Q}}{t}}\,,
\end{equation*}
that is even then the error is generically of order $t^{-1}$ only.\\
Thus \eqref{eq.straight-deltaT=infty} can be true for general wave functions only if the real
velocity $\vpsi$ converges to the asymptotic velocity $v_\infty$ in such a way that $\vpsi-v_\infty$
still goes to zero when integrated over in time. Up to now we have however no means to prove anything like that.
\end{rem}

\noindent In the second part of the paper, we consider more general wave functions than pure scattering wave functions.
Besides the scattering part $\psac[0]\in\C\subset\mathcal{H}_{\rm ac}(H)$ we will allow the wave function
$\ps[0]$
to have a bound part $\pspp[0]\in\D\subset\mathcal{H}_{\rm pp}(H)$, where $\D$ is defined as follows.\\

\begin{defi}
\label{def.pspp-less-q^(-3)}
Let $\alpha>0$. $f:\real^3\rightarrow\mathbb{C}$ is in $\D$ if
\begin{gather*}
f\in\mathcal{H}_{\rm pp}(H)\cap\ro C^\infty(H)\\
\intertext{and there exist $R>0$ and $C<\infty$ such that}
\sup\limits_{t\in\real}|e^{-iHt}f(q)|\leq C|q|^{-\nhalf[3]-\alpha}\qquad\text{and}\qquad
\sup\limits_{t\in\real}|\nabla e^{-iHt}f(q)|\leq C|q|^{-\nhalf[3]-\alpha}
\end{gather*}
for all $|q|>R$.
\end{defi}

\begin{rem}
\label{rem.why-A4}
$\pspp[0]\in\D$ ($\alpha>0$) seems to be a reasonable assumption. Indeed there is a huge
amount of
literature on the exponential decay of eigenfunctions of Schr\"{o}dinger operators, although results for the gradient
of eigenfunctions are rather rare (see \cite{simon:82, simon:00} for an overview). We wish to recall here two results on
eigenfunctions
$u\in L^2(\real^3)$, i.e. solutions of $Hu=Eu$ with $H$ as above and $E<0$\footnote{
Clearly for $V\in(V)_4$ there are no positive eigenvalues (see e.g. \cite{ikebe:60}).}.
\begin{enumerate}
\item
There exist $R>0$ and $C<\infty$ such that
\begin{equation*}
\sup\limits_{t\in\real}|e^{-iHt}u(q)|=\sup\limits_{t\in\real}|e^{-iEt}u(q)|=|u(q)|\leq C|q|^{-1}e^{-|E|^{\nhalf[1]}|q|}
\end{equation*}
for all $|q|\geq R$ (see e.g. \cite{agmon:85}).
\item
If in addition to the above $V\in K^{(1)}_3$ (where we use the notation of \cite{simon:82}, p. 467), i.e. if the
singularities of $V$ are not too bad, $u\in C^1(\Omega)$ and for every $q_0\in\Omega$
\begin{equation*}
\sup\limits_{\{q\in\Omega\mid|q_0-q|\leq 1\}}
|\nabla u(q)|\leq C\int\limits_{|q_0-q|\leq 2}|u(q)|\,dq
\end{equation*}
for some (possibly E-dependent) positive constant $C$ (q.v. \cite{simon:82}: Theorems C.2.4. and C.2.5.).
Using (i), we particulary get
$\sup\limits_{t\in\real}|\nabla e^{-iHt}u(q)|=\mathcal{O}(|q|e^{-|E|^{\nhalf[1]}|q|})$ for $|q|\rightarrow\infty$.
\end{enumerate}
\end{rem}
\hfill\\
\hfill\\
\noindent For $\ps[0]=\psac[0]+\pspp[0]$ with $\psac[0]\in\C$ and $\pspp[0]\in\D$ (for any $\alpha>0$)
we show that
the limit $v_\infty(\Qo):=\lim\limits_{t\rightarrow\infty}\qt$ still exists for $\PP^{\ps[0]}$-almost all
trajectories $\Q$. As described in the introduction we obtain that $\PP^{\ps[0]}$-almost all trajectories
are either such that their long
time behavior is governed solely by the scattering part $\psac[0]$ of the wave function, i.e. they are scattering
trajectories, or such that their long time behavior is governed solely by the bound part $\pspp[0]$ of the wave
function, i.e. they are bound trajectories. Moreover we obtain that the asymptotic velocity of a scattering trajectory
is equal to $v_\infty$, i.e. it is equal to that of a straight path.
Finally the probability distribution of $v_\infty$ has density
$|\psat(\cdot)|^2+\|\pspp[0]\|^2\delta^3(\cdot)$, that is $v_\infty$ has the same probability distribution
as in the case of a pure scattering wave function  --  except at $v_\infty=0$ where the mass
$\|\pspp[0]\|^2$ of the bound trajectories is located. These assertions are collected in\\

\begin{thm}
\label{thm.mixed}
Let $H=H_0+V$ with $V\in(V)_4$ and let zero be neither an eigenvalue nor a resonance of $H$. Let
$\ps[0]=\psac[0]+\pspp[0]$ with $\psac[0]\in\C$ and $\pspp[0]\in\D$ (for some $\alpha>0$). Then
\begin{enumerate}
\item
the Bohmian trajectories $\Q$ exist globally in time for $\PP^{\ps[0]}$-almost all initial configurations $\Qo\in\real^3$.
\item
$v_\infty(\Qo):=\lim\limits_{t\rightarrow\infty}\qt$ exists for $\PP^{\ps[0]}$-almost all $\Qo\in\real^3$ and its
probability distribution has density
$|\psat(\cdot)|^2+\|\pspp[0]\|^2\delta^3(\cdot)$, i.e. for every measurable set $\ro{A}\subset\real^3$
\begin{equation}
\label{eq.v-infty-distr-mixed}
\PP^{\ps[0]}\left( v_\infty\in\ro{A}\right)=
\begin{cases}
\int\limits_{\ro{A}}|\psat(k)|^2\,d^3k+\|\pspp[0]\|^2 &\text{if $0\in\ro{A}$},\\
\int\limits_{\ro{A}}|\psat(k)|^2\,d^3k                 &\text{if $0\not\in\ro{A}$}\,.
\end{cases}
\end{equation}
\item
$\PP^{\ps[0]}$-almost all Bohmian trajectories are either bound trajectories or scattering trajectories
and for scattering trajectories the asymptotic velocity is given by $v_{\infty}$,
i.e. for all $\eps>0$ and all $0<\gamma<2\alpha$ there exist $R>0\,,\,T>0$ and $C<\infty$ such that
\begin{gather}
\label{eq.P(bound-in-slow-ball)-mixed}
\bigg|\PP^{\ps[0]}\bigg(\bigg\{\Qo\in\real^3\mid|\Q|\leq R\big(\frac{t}{T}\big)^{\frac{1}{1+\gamma}}
        \quad\forall t\geq T\bigg\}\bigg)-\|\pspp[0]\|^2\bigg|<\eps\\
\intertext{and for $\beta:=\min\left\{\alpha,\nhalf[1]\right\}$}
\label{eq.P(v-well-behaved)-mixed}
\begin{aligned}
\bigg|\PP^{\ps[0]}\bigg(\bigg\{\Qo\in\real^3\mid|\Q|>R\frac{t}{T}\:
        \wedge\:\left|\vpsi\left(\Q,t\right)-v_{\infty}(q)\right|<C&t^{-\beta}\quad\forall t\geq T\bigg\}\bigg)-\\
&-\|\psac[0]\|^2\bigg|<\eps.
\end{aligned}
\end{gather}
\end{enumerate}
\end{thm}

\noindent We rewrite our results in terms of the trajectories.

\begin{cor}
\label{cor.straight-mixed}
Let $H=H_0+V$ with $V\in(V)_4$ and let zero be neither an eigenvalue nor a resonance of $H$. Let
$\ps[0]=\psac[0]+\pspp[0]$ with $\psac[0]\in\C$ and $\pspp[0]\in\D$ (for some $\alpha>0$).
Let $g(\Qo,T',t):=\Q[T']+v_\infty(\Qo)(t-T')$ be the straight path of a particle with the uniform velocity
$v_\infty(\Qo):=\lim\limits_{t\rightarrow\infty}\qt$.
Then for all $\eps>0$, $\delta>0$ and $\deltat>0$ there exists some $T>0$ such that
\begin{equation}
\label{eq.P(ps-straight)}
\bigg|\PP^{\ps[0]}\bigg(\bigg\{\Qo\in\real^3\mid \sup\limits_{T'\geq T}\;\sup\limits_{t\in[T',T'+\deltat]}
        \left|\Q-g(\Qo,T',t)\right|<\delta\bigg\}\bigg)-\|\psac[0]\|^2\bigg|<\eps.
\end{equation}
\end{cor}
\hfill\\

%-----------------------------------------------------------------------------------------------------------------

\section{Proof}
\label{sec.proof}
\subsection{Three Preparatory Lemmata}
\label{subsec.prep-lem}

Since the proof will mostly use properties of the Fourier transform $\psat$ of the outgoing asymptote rather
than properties of the scattering (part of the) wave function $\psac$ we give the following mapping lemma.
\begin{lem}
\label{lem.mapping}
Let $H=H_0+V$ with $V\in(V)_4$ and let zero be neither an eigenvalue nor a resonance of H. Define
$\hat{\C}$ as follows:\\
Let $g:\real^3\backslash\{0\}\rightarrow\mathbb{C}$. We say $g\in\hat{\C}$ if there is some $C<\infty$ such that
\begin{alignat*}{2}
|g(k)|&                                         \leq C\langle k\rangle^{-5}\,,&\\
|\partial^\eta_k g(k)|&                         \leq C,&|\eta|=1\,,\\
|\kappa\partial^\eta_k g(k)|&                   \leq C\langle k\rangle^{-1}\,,\qquad&|\eta|=2\,,\\
\big|\frac{\partial}{\partial|k|}g(k)\big|&     \leq C\langle k\rangle^{-5}\,,&\\
\big|\frac{\partial^2}{\partial|k|^2}g(k)\big|& \leq C\langle k\rangle^{-2}\,,&
\end{alignat*}
where $\langle\;k\;\rangle:=\left(1+k^2\right)^{\nhalf[1]}\,,\kappa=\frac{|k|}{\langle k\rangle}$ and
$\eta$ is a multi-index.\\
Then
\begin{equation*}
\ps[0]\in\C\quad\Rightarrow\quad\psat\in\hat{\C}.
\end{equation*}
\end{lem}
\hfill\\
\noindent The proof of Lemma \ref{lem.mapping} is analog to that of Lemma 3 in \cite{duerr1:04}.\\

\noindent For the proof of both Theorem \ref{thm.pure-scattering} and Theorem \ref{thm.mixed} we need (pointwise)
estimates on how fast the scattering (part of the) wave function tends to the
local plane wave $\phi_1=(it)^{-\nhalf}\exp\left(i\frac{q^2}{2t}\right)\psat\left(\xt\right)$ described in
the introduction. Since we are mainly interested in the velocity field
$\vpsi=\Im\left(\frac{\nabla\ps[]}{\ps[]}\right)$ we also need estimates
on the gradient.
\begin{lem}
\label{lem.est-psac-phi3}
Let $H=H_0+V$ with $V\in(V)_4$ and let zero be neither an eigenvalue nor a resonance of $H$. Let
$\ps[0]\in\C$. From $\ps$ we split off its (freely evolving) outgoing asymptote $\psout=e^{-iH_0t}\psout[0]$,
\begin{equation}
\label{eq.psac-psout-phi3}
\ps(q)=:\psout(q)+\phi_3(q,t)\,.
\end{equation}
Then $\ps$ tends to $\psout$ in the sense that there is some $R>0$ such that
for all $T>0$ there exists some $C_T<\infty$ such that
\begin{gather}
\label{eq.est-phi3}
|\phi_3(q,t)|\leq \frac{C_T}{|q|(t+|q|)}\qquad\forall |q|>0,
\addtocounter{equation}{1}
\tag{\theequation a}\\
\label{eq.est-Delta-phi3}
\left|\nabla\phi_3(q,t)\right|\leq \frac{C_T}{|q|(t+|q|)}\qquad\forall |q|>R
\tag{\theequation b}
\end{gather}
for all $t\geq T$.
\end{lem}
\hfill\\

\noindent The proof can be found in \cite{teufel2:99}. We give more detailed information in the appendix.\\
\begin{lem}
\label{lem.est-psac-phi1}
Let $H=H_0+V$ with $V\in(V)_4$ and let zero be neither an eigenvalue nor a resonance of $H$. Let
$\ps[0]\in\C$. From the (freely evolving) outgoing asymptote $\psout=e^{-iH_0t}\psout[0]$ of $\ps$ we split
off the local plane wave $\phi_1$,
\begin{gather}
\label{eq.psout-phi1-phi2}
\begin{split}
\psout(q)&=(it)^{-\nhalf}e^{\frac{iq^2}{2t}}\psat\left(\xt\right)+
        (2\pi it)^{-\nhalf}e^{\frac{iq^2}{2t}}\int\limits_{\real^3}e^{-i\frac{q\cdot y}{t}}
        \left(e^{\frac{iq^2}{2t}}-1\right)\psout[0](y)\,d^3y=\\
&=(it)^{-\nhalf}e^{\frac{iq^2}{2t}}\psat\left(\xt\right)+(2\pi)^{-\nhalf}\int\limits_{\real^3}e^{i(k\cdot q-\nhalf[k^2t])}
        \left(\psat(k)-\psat\left(\xt\right)\right)\,d^3k=\\
&=:\phi_1(q,t)+\phi_2(q,t)\,.
\end{split}\\
\intertext{Then there is some $C<\infty$ such that}
\label{eq.est-phi2}
\left|\phi_2(q,t)\right|\leq Ct^{-2},
\addtocounter{equation}{1}
\tag{\theequation a}\\
\label{eq.est-Delta-phi2}
\left|\nabla\phi_2(q,t)+(it)^{-\nhalf}e^{\frac{iq^2}{2t}}\nabla\psat\left(\xt\right)\right|=
    \left|\nabla\psout(q)-i\xt\phi_1(q,t)\right|\leq Ct^{-2}
\tag{\theequation b}
\end{gather}
for all $q\in\real^3$ and $t\not=0$.\\
Furthermore,
\begin{equation}
\label{eq.est-psac-l2}
\lim\limits_{t\rightarrow\infty}\|\ps-\phi_1(\cdot,t)\|=0\,.
\end{equation}
\end{lem}
\hfill\\

\noindent The proof of the pointwise estimates \eqref{eq.est-phi2} and \eqref{eq.est-Delta-phi2} can be found
in \cite{duerr1:04}. Also \eqref{eq.est-psac-l2} is a standard result.
We give more detailed information in the appendix.\\

\begin{rem}
\label{rem.why-C}
The estimates \eqref{eq.est-phi3} and \eqref{eq.est-Delta-phi3} resp. \eqref{eq.est-phi2}
and \eqref{eq.est-Delta-phi2} were derived by Teufel, D\"urr and M\"unch-Berndl \cite{teufel2:99} resp.
D\"urr, Moser and Pickl \cite{duerr1:04} using generalized eigenfunctions.
It is the properties of the eigenfunctions, which concerning smoothness and boundedness are rather poor in
general, that dictate the form of $\hat{\C}$ and thus also (through Lemma \ref{lem.mapping})
the overall form of our conditions on the scattering (part of the) wave function, i.e. of $\C$.
For a discussion of how this comes about see \cite{duerr1:04}.
\end{rem}

%---------------------------------------------------------------------------------------------------------------

\subsection{Proof of Theorem \ref{thm.pure-scattering} and Corollary \ref{cor.straight-pure}}
\label{subsec.thm1-cor1}

\begin{pro}[of Theorem \ref{thm.pure-scattering}]
$(i)$ is a direct consequence of Corollary 3.2 in \cite{berndl:95} resp. Corollary 4 in \cite{teufel:04}.\\

\noindent For technical reasons we continue with the proof of $(iii)$.\\
For $\delta_1>0$, $\delta_2>0$ and $b>a>0$ we define the sets
\begin{gather}
\label{eq.def-setonea}
\setonea:=\left\{k\in\real^3\mid|\psat(k)|>\delta_1\;\wedge\; a<|k|<b\right\}\\
\intertext{and inner subsets thereof:}
\label{eq.def-settwoa}
\settwoa:=\left\{k\in\real^3\mid\U(k)\subset\setonea\right\}
\end{gather}
where $\U(k_0)=\{k\in\real^3\mid |k-k_0|<\delta_2\}$ is the open ball around $k_0$ with radius $\delta_2$.
We show that
for all $\delta_1>0\,,\:\delta_2>0$ and $b>a>0$ there exists some $T>0$ and some suitable $C<\infty$ (depending on
$\delta_1\,,a$ and $b$) such that
\begin{equation}
\label{eq.control-v-pure}
\qtp[T]\in\settwoa\qquad\Rightarrow\qquad\left|\vpsi\left(\Qp,t\right)-\qtp\right|<\frac{C}{3}t^{-\nhalf[1]}\quad
\forall t\geq T\,.
\end{equation}
For this we first show that there is some $\tilde{C}<\infty$ such that
\begin{equation}
\label{eq.est-v(q)-pure}
\left|\vpsi(q,t)-\xt\right|\leq\frac{\tilde{C}t^{-2}}{\left|\ps(q)\right|}\left(1+\frac{|q|}{t}\right)
        \left(1+\tx\right)
\end{equation}
for $t$ and $|q|$ big enough and such that $\ps(q)\not=0$.\\
By \eqref{eq.eq-of-motion} we have for $q\in\real^3$ and $t\in\real$ such that $\ps(q)\not=0$
\begin{equation*}
\left|\vpsi(q,t)-\xt\right|=\left|\Im\left(\frac{\nabla\ps(q)}{\ps(q)}-i\xt\right)\right|
\leq|\ps(q)|^{-1}\left|\nabla\ps(q)-i\xt\ps(q)\right|.
\end{equation*}
To estimate $\left|\nabla\ps(q)-i\xt\ps(q)\right|$ for $t$ and $|q|$ big enough we use Lemma \ref{lem.est-psac-phi3}
and Lemma \ref{lem.est-psac-phi1} and get for some suitable $\tilde{C}<\infty$
\begin{equation*}
\begin{split}
\bigg|\nabla\ps(q)-&i\xt\ps(q)\bigg|
        \leq\left|\nabla\ps(q)-i\xt\phi_1(q,t)\right|+\frac{|q|}{t}\left|\ps(q)-\phi_1(q,t)\right|\leq\\
    &\leq\left|\nabla\psout(q)-i\xt\phi_1(q,t)\right|+\left|\nabla\phi_3(q,t)\right|+
        \frac{|q|}{t}\left(\left|\psout(q)-\phi_1(q,t)\right|+\left|\phi_3(q,t)\right|\right)\leq\\
    &\leq \tilde{C}\left(t^{-2}+\frac{1}{|q|(t+|q|)}\right)\left(1+\frac{|q|}{t}\right)
        \leq \tilde{C}t^{-2}\left(1+\tx\right)\left(1+\frac{|q|}{t}\right).
\end{split}
\end{equation*}
From this \eqref{eq.est-v(q)-pure} follows.\\
Now let $\delta_1>0\,,\,\delta_2>0$ and $b>a>0$.
To get \eqref{eq.control-v-pure} we shall show that for $T$ big enough
\begin{equation}
\label{eq.settwoa-good-pure}
\qtp[T]\in\settwoa\quad\mbox{implies}\quad\qtp\in\U\left(\qtp[T]\right)\subset\setonea\quad\forall t\geq T
\end{equation}
and that there is some $C<\infty$ (depending on $\delta_1\,,a$ and $b$) such that for all $t\geq T$
\begin{equation}
\label{eq.setonea-good-pure}
\xt\in\setonea\quad\mbox{implies}\quad \left|\vpsi(q,t)-\xt\right|<\frac{C}{3}t^{-\nhalf[1]}\,.
\end{equation}
We start with \eqref{eq.setonea-good-pure}.\\
Let $\xt\in\setonea$. Then by Lemma \ref{lem.est-psac-phi3} and Lemma \ref{lem.est-psac-phi1}
\begin{align*}
t^{\nhalf}|\ps(q)|&\geq t^{\nhalf}\big[|\phi_1(q,t)|-|\phi_2(q,t)|-|\phi_3(q,t)|\big]>\\
&>\left|\psat\left(\xt\right)\right|-C'\left(t^{-\nhalf[1]}+t^{\nhalf}\frac{1}{|q|(t+|q|)}\right)\geq
        \left|\psat\left(\xt\right)\right|-C't^{-\nhalf[1]}\left(1+\tx\right)
\end{align*}
for some suitable $C'<\infty$ and $t$ big enough. In the last step we used
\begin{equation*}
\frac{1}{|q|(t+|q|)}\leq\frac{1}{|q|t}=t^{-2}\tx\,.
\end{equation*}
Since by \eqref{eq.def-setonea} $\left|\psat\left(\xt\right)\right|>\delta_1$ and $\tx<a$, for $t$
big enough we can choose $C't^{-\nhalf[1]}\left(1+\tx\right)\leq\nhalf[\delta_1]$ and obtain
\begin{equation*}
t^{\nhalf}|\ps(q)|\geq\nhalf[\delta_1]\,.
\end{equation*}
Using this, $\frac{|q|}{t}<b$ and again $\tx<a$ we get by \eqref{eq.est-v(q)-pure}
\begin{equation*}
\left|\vpsi(q,t)-\xt\right|\leq \frac{C}{3}t^{-\nhalf[1]}
\end{equation*}
for some ($\delta_1\,,a$ and $b$ dependent) $C<\infty$ and for $t$ big enough.\\
Now let $\qtp[T]\in\settwoa$. Suppose that there exists some $t_1>T$ such that $\qtp[t_1]\not\in\U\left(\qtp[T]\right)$.
Since $\Qp$ is continuous in $t$ (it is a solution of the first order ODE \eqref{eq.eq-of-motion} that
exists globally in time) this implies that the first exit time
\begin{equation*}
t_{\rm ex}(q):=\max\left\{s>T\mid \qtp[s]\not\in\U\left(\qtp[T]\right)\;\wedge
        \;\qtp[\tau]\in\U\left(\qtp[T]\right)\quad\forall\, T\leq \tau<s\right\}
\end{equation*}
exists and that
$\left|\qtp[t_{\rm ex}(q)]-\qtp[T]\right|=\delta_2$. However,
$\qtp[\tau]\in\U\left(\qtp[T]\right)\subset\setonea$ for all $T\leq\tau<t_{\rm ex}(q)$, i.e. by
\eqref{eq.setonea-good-pure} we have for $T$ big enough
\begin{equation}
\label{eq.Q-Cauchy-pure}
\begin{split}
\left|\qtp[t_{\rm ex}(q)]-\qtp[T]\right|&=
        \int\limits_{T}^{t_{\rm ex}(q)}\left|\frac{\partial}{\partial\tau}\qtp[\tau]\right|\,d\tau
        \leq\int\limits_{T}^{t_{\rm ex}(q)}\frac{1}{\tau}\left|\vpsi\left(\Qp[\tau],\tau\right)
                -\qtp[\tau]\right|\,d\tau <\\
        &<\int\limits_{T}^{\infty}\frac{C}{3\tau}\tau^{-\nhalf[1]}\,d\tau = \frac{2}{3}CT^{-\nhalf[1]}<\delta_2.
\end{split}
\end{equation}
Since this is a contradiction \eqref{eq.settwoa-good-pure} holds. Using \eqref{eq.setonea-good-pure} (with
$q$ replaced by $\Qp$) we obtain
\begin{equation*}
\qtp[T]\in\settwoa\;\Rightarrow\;
\qtp\in\setonea\;\forall t\geq T\;\Rightarrow\;
\left|\vpsi\left(\Qp,t\right)-\qtp\right|<\frac{C}{3}t^{-\nhalf[1]}\;\forall t\geq T\,,
\end{equation*}
that is we obtain \eqref{eq.control-v-pure}.\\

\noindent Next we show that there is a measurable set $\G$ of "good"
initial configurations $q$ for which the velocity $\vpsi\left(\Qp,t\right)$ is well behaved:
\begin{equation}
\label{eq.v-well-behaved-pure}
\begin{split}
&v_\infty(q):=\lim\limits_{t\rightarrow\infty}\qtp\qquad\mbox{exists and}\\
&\left|\vpsi(\Qp,t)-v_{\infty}(q)\right|\leq Ct^{-\nhalf[1]}\qquad\forall t\geq T\,.
\end{split}
\end{equation}
Indeed, with the help of \eqref{eq.control-v-pure} we can rewrite \eqref{eq.Q-Cauchy-pure} to get
\begin{equation*}
\qtp[T]\in\settwoa\qquad\Rightarrow\qquad\left|\qtp[t_1]-\qtp[t_2]\right|<\frac{2}{3}Ct_1^{-\nhalf[1]}
        \quad\forall\,t_2\geq t_1\geq T\,.
\end{equation*}
Thus $\left(\qtp\right)$ is a Cauchy sequence and $v_\infty(q):=\lim\limits_{t\rightarrow\infty}\qtp$
exists whenever $\qtp[T]\in\settwoa$ for some $\delta_1>0\,,\,\delta_2>0\,,\,b>a>0$ and for $T$ big enough.
Then also (again using \eqref{eq.control-v-pure})
\begin{align*}
\left|\vpsi\left(\Qp,t\right)-v_\infty(q)\right|&\leq
        \left|\vpsi\left(\Qp,t\right)-\qtp\right|+\lim\limits_{s\rightarrow\infty}\left|\qtp-\qtp[s]\right|<\\
    &<Ct^{-\nhalf[1]}
\end{align*}
for all $t\geq T$. Thus we have shown that \eqref{eq.v-well-behaved-pure} holds for $q\in\G$ with
\begin{equation*}
\G:=\left\{q\in\real^3\,\big\vert\,\qtp[T]\in\settwoa\right\}\,,
\end{equation*}
where $T>0$ was big enough and $\delta_1>0\,,\,\delta_2>0$ and $b>a>0$ were still arbitrary.\\

\noindent Next we show that we can adjust $\delta_1\,,\,\delta_2\,,b\,,a$ and $T$ in
such a way that the set $\G$ of "good" initial configurations has (nearly) full measure. Note that this especially
implies almost sure existence of $v_\infty$.\\
Let $\eps>0$. We show that for $\delta_1\,,\,\delta_2$ and $a$ small and $b$ and $T$
big enough
\begin{equation}
\label{eq.P(not-in-B)-pure}
1-\PP^{\ps[0]}\left(\G\right)=\PP^{\ps[0]}\left(\left\{\Qo\in\real^3\,\big\vert\,\qt[T]\not\in\settwoa\right\}\right)
    <\eps\,.
\end{equation}
Since by \eqref{eq.v-well-behaved-pure}
\begin{equation*}
\G\subset\left\{q\in\real^3\mid
    \left|\vpsi\left(\Qp,t\right)-v_{\infty}(q )\right|<Ct^{-\nhalf[1]}\quad\forall\,t\geq T\right\}
\end{equation*}
this then gives us \eqref{eq.P(v-well-behaved)-pure}:
\begin{equation*}
\PP^{\ps[0]}\left(\left\{\Qo\in\real^3\mid
    \left|\vpsi\left(\Q,t\right)-v_{\infty}(\Qo)\right|<Ct^{-\nhalf[1]}\quad\forall\,t\geq T\right\}\right)
    \geq\PP^{\ps[0]}(\G)>1-\eps\,.
\end{equation*}
Back to the proof of \eqref{eq.P(not-in-B)-pure}.
Using equivariance \eqref{eq.equivariance} and
\begin{equation*}
\settwoa^{\rm c}=\setonea^{\rm c}+\left\{k\in\real^3\mid k\in\setonea\quad\wedge\quad\U(k)\not\subset\setonea\right\}
\end{equation*}
we get (for simplicity we write $q$ instead of $\Qo$)
\begin{equation}
\label{eq.qt-not-in-settwoa-split}
\begin{split}
\PP^{\ps[0]}&\left(\left\{q\in\real^3\mid\qtp[T]\not\in\settwoa\right\}\right)=
        \PP^{\ps[T]}\left(\left\{q\in\real^3\mid\xt[T]\not\in\settwoa\right\}\right)\leq\\
&\leq\PP^{\ps[T]}\left(\left\{q\in\real^3\mid\xt[T]\not\in\setonea\right\}\right)+
        \PP^{\ps[T]}\left(\left\{q\in\real^3\mid\xt[T]\in\setonea\:\wedge
                \:\U(\xt[T])\not\subset\setonea\right\}\right)\,.
\end{split}
\end{equation}
To estimate the second term in the last line we use that $\psat(k)$ is continuous (outside $k=0$)
by Lemma \ref{lem.mapping}. Then $\setonea$ is open (recall \eqref{eq.def-setonea}) and thus
\begin{equation*}
\left\{q\in\real^3\mid\xt[T]\in\setonea\:\wedge\:\U(\xt[T])\not\subset\setonea\right\}\rightarrow \emptyset
\end{equation*}
as $\delta_2\rightarrow 0$. Therefore
\begin{equation*}
\lim\limits_{\delta_2\rightarrow 0}\PP^{\ps[T]}\left(\left\{q\in\real^3\mid\xt[T]\in\setonea\:\wedge
    \:\U(\xt[T])\not\subset\setonea\right\}\right)=0\,,
\end{equation*}
that is the second term can be made smaller than $\nhalf[\eps]$ for $\delta_2$ small enough and we are left with the task
to provide an estimate for the first term in the last line of \eqref{eq.qt-not-in-settwoa-split}.\\
For convenience define
\begin{equation*}
\setonea[C](T):=\left\{q\in\real^3\mid\xt[T]\not\in\setonea\right\}\,.
\end{equation*}
Then
\begin{align*}
\PP^{\ps[T]}\left(\left\{q\in\real^3\mid\xt[T]\not\in\setonea\right\}\right)&=
        \|\chi_{\setonea[C](T)}\ps[T]\|^2\leq\\
&\leq\left(\|\chi_{\setonea[C](T)}\phi_1(\cdot,T)\|+\left\|\ps[T]-\phi_1(\cdot,T)\right\|\right)^2\,.
\end{align*}
By \eqref{eq.est-psac-l2} the second term can be made arbitrary small for $T$ big enough. For the
first term we substitute $k:=\xt[T]$ and get (with $\phi_1(q,t)=(it)^{-\nhalf}
\exp\left(i\frac{q^2}{2t}\right)\psat\left(\xt\right)$ the local plane wave defined in Lemma \ref{lem.est-psac-phi1})
\begin{align*}
\big\|\chi&_{\setonea[C](T)}\phi_1(\cdot,T)\big\|^2=\int\limits_{\setonea[C](T)}T^{-3}\left|\psat\left(
        \xt[T]\right)\right|^2\,d^3q=\int\limits_{\setonea^{\rm c}}|\psat(k)|^2\,d^3k=\\
&=\int\limits_{|\psat(k)|\leq\delta_1}|\psat(k)|^2\,d^3k+
        \int\limits_{|k|\leq a}|\psat(k)|^2\,d^3k+
        \int\limits_{|k|\geq b}|\psat(k)|^2\,d^3k\leq\\
&\leq\int\limits_{|\psat(k)|\leq\delta_1\wedge|k|<b}|\psat(k)|^2\,d^3k+
        \int\limits_{|k|\leq a}|\psat(k)|^2\,d^3k+
        2\int\limits_{|k|\geq b}|\psat(k)|^2\,d^3k\leq\\
&\leq\frac{4}{3}\pi b^3{\delta_1}^2+\frac{4}{3}\pi a^3\sup\limits_{k\in\real^3\backslash\{0\}}|\psat(k)|^2
        +2\int\limits_{|k|\geq b}|\psat(k)|^2\,d^3k\,.
\end{align*}
Since $\psat$ is square integrable the third term can be made arbitrary small for $b$ big enough. Then the
first term can be diminished at will by decreasing $\delta_1$ accordingly. By Lemma \ref{lem.mapping}
$\sup\limits_{k\in\real^3\backslash\{0\}}|\psat(k)|$ is bounded, so the second term can be made arbitrary small for $a$
small enough. Thus we have shown \eqref{eq.P(not-in-B)-pure}.\\
\\
\noindent Finally we prove $(ii)$.\\
Since we have already shown that $v_\infty$ exists for almost all initial conditions $q_0\in\real^3$ it is only
left to show that $v_\infty$ is $|\psat|^2$-distributed. Let $A\subset\real^3$ be measurable. By
dominated convergence and equivariance \eqref{eq.equivariance}
\begin{align*}
\PP^{\ps[0]}\left(v_\infty\in A\right)&=
        \lim\limits_{t\rightarrow\infty}\PP^{\ps[0]}\big(\{q\in\real^3\mid\;\qtp\in A\}\big)=\\
&=\lim\limits_{t\rightarrow\infty}\PP^{\ps}\big(\{q\in\real^3\mid\;\xt\in A\}\big)=
        \lim\limits_{t\rightarrow\infty}\|\chi_{\xt\in A}\ps\|^2\,.
\end{align*}
By \eqref{eq.est-psac-l2} this yields (again with $k:=\xt$)
\begin{align*}
\PP^{\ps[0]}\left(v_\infty\in A\right)&=
        \lim\limits_{t\rightarrow\infty}\|\chi_{\xt\in A}\phi_1(\cdot,t)\|^2=
        \lim\limits_{t\rightarrow\infty}\int\limits_{\xt\in A}t^{-3}\left|\psat\left(\xt\right)
        \right|^2\,d^3q=\\
&=\int\limits_{A}\left|\psat(k)\right|^2\,d^3k\,.
\end{align*}
\end{pro}
\hfill\\

\begin{pro}[of Corollary \ref{cor.straight-pure}]
Let $\eps>0\,,\,\delta>0$ and $\deltat>0$. By Theorem \ref{thm.pure-scattering} $(iii)$
there exists some
$T>0$ and some $C<\infty$ such that
\begin{equation*}
\PP^{\ps[0]}\left(\left\{\Qo\in\real^3\mid\left|\vpsi\left(\Q,t\right)-
        v_{\infty}(\Qo)\right|<Ct^{-\nhalf[1]}\quad\forall t\geq T\right\}\right)>1-\eps\,.
\end{equation*}
But then also
\begin{equation*}
\PP^{\ps[0]}\left(\left\{\Qo\in\real^3\mid\sup\limits_{t\geq T}\left|\vpsi\left(\Q,t\right)-
        v_{\infty}(\Qo)\right|<\frac{\delta}{\deltat}\right\}\right)>1-\eps
\end{equation*}
if $T$ is big enough.\\
Now let $\Qo\in\real^3$ be such that $\sup\limits_{t\geq T}\left|\vpsi\left(\Q,t\right)-
v_{\infty}(\Qo)\right|<\frac{\delta}{\deltat}$ and let $T'\geq T$. Then
\begin{equation*}
\big|\Q-\left[\Q[T']+v_\infty(\Qo)(t-T')\right]\big|=
\left|\int\limits_{T'}^{t}\left(\vpsi\left(\Q[\tau],\tau\right)-v_\infty(\Qo)\right)\,d\tau\right|<
\frac{\delta}{\deltat}(t-T')\leq\delta
\end{equation*}
for all $t\in[T',T'+\deltat]$. So we get \eqref{eq.P(psac-straight)}.
\end{pro}
\hfill\\

\subsection{Proof of Theorem \ref{thm.mixed} and Corollary \ref{cor.straight-mixed}}
\label{subsec.thm2-cor2}

Let $\ps[0]=\psac[0]+\pspp[0]$. We need that the support of the
scattering part $\psac$ and that of the bound part $\pspp$ gets separated for big times. As described
in the introduction and shown in Subsection \ref{subsec.prep-lem} (Lemma \ref{lem.est-psac-phi1},
Equation \eqref{eq.est-psac-l2}) the support of $\psac$ moves out to spatial infinity, so we are done if we can show that
the support of $\pspp$ stays concentrated around the scattering center for all times, i.e. if we can show that
for all $\eps>0$ there exists some $R>0$ such that
\begin{equation}
\label{eq.est-pspp-l2}
\sup_{t\in\real}\int\limits_{B^{\rm c}_R}|\pspp(q)|^2\,d^3q<\eps.
\end{equation}
But that is a well known feature of bound wave functions (see e.g. \cite{ruelle:69} or \cite{perry:83}
Theorem 2.1 and Example 2.2).\\

\begin{pro}[of Theorem \ref{thm.mixed}]
Again $(i)$ is a direct consequence of Corollary 3.2 in \cite{berndl:95} resp. Corollary 4 in \cite{teufel:04}.\\

\noindent We start with the proof of $(iii)$.\\
Let $\setonea$ and $\settwoa$ be the sets defined in the proof of Theorem \ref{thm.pure-scattering}
(Equations \eqref{eq.def-setonea} and \eqref{eq.def-settwoa} respectively) and let $\beta=\min\{\alpha,\nhalf[1]\}$.
We show that for all $\delta_1>0\,,\:\delta_2>0$ and $b>a>0$ there exists some $T>0$ such that
\begin{equation}
\label{eq.control-v-mixed}
\qtp[T]\in\settwoa\qquad\Rightarrow\qquad\left|\Qp\right|>at\quad\wedge\quad
        \left|\vpsi\left(\Qp,t\right)-\qtp\right|<\frac{C}{3}t^{-\beta}
\end{equation}
for all $t\geq T$ and some suitable $C<\infty$ (depending on $\delta_1,\, a$ and $b$).\\
Analog to \eqref{eq.est-v(q)-pure} in the proof of Theorem \ref{thm.pure-scattering} we start by showing
\begin{equation}
\label{eq.est-v(q)-mixed}
\left|\vpsi(q,t)-\xt\right|\leq\frac{\tilde{C}t^{-\nhalf-\beta}}{\left|\ps(q)\right|}\left(1+\frac{|q|}{t}\right)
        \left[\left(1+\tx\right)+\left(\tx\right)^{\nhalf+\alpha}\right]
\end{equation}
for some suitable $\tilde{C}$ and for all $t$ and $|q|$ big enough such that $\ps(q)\not=0$.\\
By \eqref{eq.eq-of-motion} we have for $q\in\real^3$ and $t\in\real$ such that $\ps(q)\not=0$
\begin{equation*}
\left|\vpsi(q,t)-\xt\right|=\left|\Im\left(\frac{\nabla\ps(q)}{\ps(q)}-i\xt\right)\right|
\leq|\ps(q)|^{-1}\left|\nabla\ps(q)-i\xt\ps(q)\right|.
\end{equation*}
Thus we need to estimate
\begin{equation*}
\left|\nabla\ps(q)-i\xt\ps(q)\right|\leq\left|\nabla\ps(q)-i\xt\phi_1(q,t)\right|+
    \frac{|q|}{t}\left|\ps(q)-\phi_1(q,t)\right|
\end{equation*}
for big $t$ and $|q|$. Here $\phi_1=(it)^{-\nhalf}\exp\left(i\frac{q^2}{2t}\right)\psat\left(\xt\right)$
is the local plane wave defined in Lemma \ref{lem.est-psac-phi1}.
Since $\pspp[0]\in\D$ and $\beta=\min\{\alpha,\nhalf[1]\}$ we have (using Lemma \ref{lem.est-psac-phi3} and Lemma
\ref{lem.est-psac-phi1} in the same way as in the proof of \eqref{eq.est-v(q)-pure})
\begin{gather*}
\begin{split}
\big|\ps(q)&-\phi_1(q,t)\big|\leq\left|\psac(q)-\phi_1(q,t)\right|+\left|\pspp(q)\right|\leq
    \left|\phi_2(q,t)\right|+\left|\phi_3(q,t)\right|+\left|\pspp(q)\right|\leq\\
&\leq C'\left[t^{-2}+\frac{1}{|q|(t+|q|)}+|q|^{-\nhalf-\alpha}\right]
        \leq C'\left[t^{-2}+\frac{1}{|q|t}+|q|^{-\nhalf-\alpha}\right]\\
&\leq C't^{-\nhalf-\beta}\left[t^{-\nhalf[1]+\beta}\left(1+\tx\right)+
        t^{-\alpha+\beta}\left(\tx\right)^{\nhalf+\alpha}\right]
        \leq\tilde{C}t^{-\nhalf-\beta}\left[\left(1+\tx\right)+\left(\tx\right)^{\nhalf+\alpha}\right]
\end{split}\\
\intertext{and}
\begin{split}
\left|\nabla\ps(q)-i\xt\phi_1(q,t)\right|&\leq\left|\nabla\psac(q)-i\xt\phi_1(q,t)\right|+\left|\nabla\pspp(q)\right|\leq\\
&\leq\left|\nabla\psout(q)-i\xt\phi_1(q,t)\right|+\left|\nabla\phi_3(q,t)\right|+\left|\nabla\pspp(q)\right|\leq\\
&\leq C'\left[t^{-2}+\frac{1}{|q|(t+|q|)}+|q|^{-\nhalf-\alpha}\right]
    \leq \tilde{C}t^{-\nhalf-\beta}\left[\left(1+\tx\right)+\left(\tx\right)^{\nhalf+\alpha}\right]
\end{split}
\end{gather*}
for some suitable $\tilde{C}<\infty$ and $t$ and $|q|$ big enough. So
\begin{equation*}
\left|\nabla\ps(q)-i\xt\ps(q)\right|\leq\tilde{C}t^{-\nhalf-\beta}\left(1+\frac{|q|}{t}\right)
        \left[\left(1+\tx\right)+\left(\tx\right)^{\nhalf+\alpha}\right]
\end{equation*}
and we get \eqref{eq.est-v(q)-mixed}.\\
Since $\qtp\in\setonea$ directly implies $\left|\Qp\right|>at$ \eqref{eq.control-v-mixed} follows from
\eqref{eq.est-v(q)-mixed} in exactly the same way as \eqref{eq.control-v-pure} followed from
\eqref{eq.est-v(q)-pure}.\\

\noindent Next we show that there is a measurable set $\G$ of "good"
initial configurations $q$ for which the velocity $\vpsi\left(\Qp,t\right)$ is well behaved in the sense that
asymptotically it is that of a straight line:
\begin{equation}
\label{eq.settwoa-good-mixed}
\begin{split}
&v_\infty(q):=\lim\limits_{t\rightarrow\infty}\qtp\quad\mbox{exists and}\\
&\left|\Qp\right|>at\quad\wedge\quad\left|\vpsi\left(\Qp,t\right)-v_\infty(q)\right|<Ct^{-\beta}\quad\forall t\geq T\,.
\end{split}
\end{equation}
We use \eqref{eq.control-v-mixed} in the same way we used
\eqref{eq.control-v-pure} in the proof of \eqref{eq.v-well-behaved-pure} of Theorem \ref{thm.pure-scattering}
to get that \eqref{eq.settwoa-good-mixed} holds for $q\in\G$ with
\begin{equation*}
\G:=\left\{q\in\real^3\,\big\vert\,\qtp[T]\in\settwoa\right\}\,,
\end{equation*}
where $T>0$ is big enough and $\delta_1>0\,,\,\delta_2>0$ and $b>a>0$ are still arbitrary.\\

\noindent Next we show \eqref{eq.P(v-well-behaved)-mixed}, that is we show that the set of initial
configurations $q_0$ for which \eqref{eq.settwoa-good-mixed} holds has measure arbitrary close to
$\|\psac[0]\|^2$. For any $\delta_1>0\,,\,\delta_2>0\,,b>a>0$ and for $T>0$ big enough we have (with $R:=aT$)
\begin{multline*}
\PP^{\ps[0]}\left(\left\{\Qo\in\real^3\mid\qt[T]>a\right\}\right)\geq\\
\geq\PP^{\ps[0]}\left(\left\{\Qo\in\real^3\mid|\Q|>R\frac{t}{T}\:
    \wedge\:\left|\vpsi\left(\Q,t\right)-v_{\infty}(q)\right|<Ct^{-\beta}\quad\forall t\geq T\right\}\right)\geq\\
\geq\PP^{\ps[0]}\left(\G\right)\,,
\end{multline*}
where the first inequality is trivial and the second follows from what we just said in \eqref{eq.settwoa-good-mixed}.
We shall show that for
any $\eps>0$ there are $\delta_1>0\,,\,\delta_2>0$ and $a>0$ small and $b>a$ and $T>0$ big enough such that
\begin{gather}
\label{eq.P(qt-in-B)>-mixed}
\PP^{\ps[0]}\left(\G\right)>\|\psac[0]\|^2-\eps
\intertext{and}
\label{eq.P(Q-bigger-aT)<-mixed}
\PP^{\ps[0]}\left(\left\{\Qo\in\real^3\mid\qt[T]>a\right\}\right)<\|\psac[0]\|^2+\eps\,.
\end{gather}
Thus
\begin{multline*}
\|\psac[0]\|^2+\eps>\\
>\PP^{\ps[0]}\left(\left\{\Qo\in\real^3\mid|\Q|>R\frac{t}{T}\:
    \wedge\:\left|\vpsi\left(\Q,t\right)-v_{\infty}(q)\right|<Ct^{-\beta}\;\forall t\geq T\right\}\right)>\\
>\|\psac[0]\|^2-\eps\,,
\end{multline*}
that is we get \eqref{eq.P(v-well-behaved)-mixed}.\\
Let $\eps>0$. First we prove \eqref{eq.P(qt-in-B)>-mixed}. By Schwarz inequality
(again we write $q$ instead of $\Qo$ for simplicity)
\begin{align*}
\PP^{\ps[0]}(\G^{\rm c})&=\PP^{\ps[0]}\left(\left\{q\in\real^3\mid\qtp[T]\not\in\settwoa\right\}\right)=
        \left\langle\psac[0]+\pspp[0]\big|\chi_{\qtp[T]\not\in\settwoa}\left(\psac[0]+\pspp[0]\right)\right\rangle\leq\\
&\leq\left\|\chi_{\qtp[T]\not\in\settwoa}\psac[0]\right\|^2+\big\|\pspp[0]\big\|^2+
        2\big\|\pspp[0]\big\|\,\left\|\chi_{\qtp[T]\not\in\settwoa}\psac[0]\right\|\,.
\end{align*}
Note however, that (with
$\tilde{\Psi}^{ac}_{0}:=\frac{\psac[0]}{\|\psac[0]\|}$ the
normalized scattering part of the wave function)
\begin{equation*}
\left\|\chi_{\qtp[T]\not\in\settwoa}\psac[0]\right\|^2=
        \|\psac[0]\|^2\;\PP^{\tilde{\Psi}^{ac}_{0}}\left(\left\{q\in\real^3\mid\qtp[T]\not\in\settwoa\right\}\right)
\end{equation*}
and that we already showed (\eqref{eq.P(not-in-B)-pure} in the proof of Theorem \ref{thm.pure-scattering})
that this can be made arbitrary
small for $\delta_1\,,\,\delta_2$ and $a$ small and $b$ and $T$ big enough.
Thus
\begin{equation*}
\PP^{\ps[0]}(\G^{\rm c})=\PP^{\ps[0]}\left(\left\{q\in\real^3\mid\qtp[T]\not\in\settwoa\right\}\right)<\|\pspp[0]\|^2+
        \eps\,.
\end{equation*}
Since $\|\pspp[0]\|^2+\|\psac[0]\|^2=1$ this gives \eqref{eq.P(qt-in-B)>-mixed}:
\begin{equation*}
\PP^{\ps[0]}\left(\G\right)=1-\PP^{\ps[0]}\left(\G^{\rm c}\right)>\|\psac[0]\|^2-\eps\,.
\end{equation*}
Next, to get \eqref{eq.P(Q-bigger-aT)<-mixed}, we note that by equivariance
\eqref{eq.equivariance} and again Schwarz inequality we also have
\begin{equation*}
\begin{split}
\PP^{\ps[0]}\bigg(\bigg\{q\in\real^3\mid\qtp[T]>a\bigg\}\bigg)&=
        \PP^{\ps[T]}\left(\left\{q\in\real^3\mid\xt[T]>a\right\}\right)\leq\\
&\leq\left\|\chi_{|q|>aT}\pspp[T]\right\|^2+\big\|\psac[0]\big\|^2+
        2\big\|\psac[0]\big\|\,\left\|\chi_{|q|>aT}\pspp[T]\right\|\,.
\end{split}
\end{equation*}
By \eqref{eq.est-pspp-l2} $\left\|\chi_{|q|>aT}\pspp[T]\right\|$
can be made arbitrary small for $T$ big enough and we obtain \eqref{eq.P(Q-bigger-aT)<-mixed}.\\

\noindent We proceed to prove \eqref{eq.P(bound-in-slow-ball)-mixed}, that is we show that the set of
trajectories moving out to spatial infinity like $t^{\frac{1}{1+\gamma}}$ has measure arbitrary close to
$\|\pspp[0]\|^2$. For this we show that for $a$ small and $T$ big enough
\begin{gather}
\label{eq.P(Q<aT)-mixed}
\left|\PP^{\ps[0]}\left(\left\{\Qo\in\real^3\mid|\Q[T]|\leq aT\right\}\right)-\|\pspp[0]\|^2\right|<\nhalf[\eps]\\
\intertext{and that for any $0<\gamma<2\alpha$ there is some $a>0$ small and some $T>0$ big enough such that}
\label{eq.P(Q<aT-and-Q>)-mixed}
\PP^{\ps[0]}\bigg(\bigg\{\Qo\in\real^3\mid|\Q[T]|\leq aT\quad\wedge\quad\exists\,t>T:\,
        |\Q|>aT\big(\frac{t}{T}\big)^{\frac{1}{1+\gamma}}\bigg\}\bigg)<\nhalf[\eps]\,.
\end{gather}
Then
\begin{gather*}
\begin{split}
\PP^{\ps[0]}\bigg(\bigg\{\Qo\in\real^3\mid|\Q|\leq aT\big(\frac{t}{T}\big)^{\frac{1}{1+\gamma}}
        \quad\forall\,t\geq T\bigg\}\bigg)&\leq
        \PP^{\ps[0]}\left(\left\{\Qo\in\real^3\mid|\Q[T]|\leq aT\right\}\right)<\\
&<\|\pspp[0]\|^2+\nhalf[\eps]
\end{split}\\
\intertext{and}
\begin{split}
\PP&^{\ps[0]}\bigg(\bigg\{\Qo\in\real^3\mid\exists\,t\geq T:\,|\Q|>aT\big(\frac{t}{T}\big)^{\frac{1}{1+\gamma}}
    \bigg\}\bigg)\leq\\
&\leq\PP^{\ps[0]}\left(\left\{\Qo\in\real^3\mid|\Q[T]|>aT\right\}\right)+\\
&\phantom{\leq\PP^{\ps[0]}(\{\Qo\in\real^3}+\PP^{\ps[0]}\bigg(\bigg\{\Qo\in\real^3\mid
        |\Q[T]|\leq aT\quad\wedge\quad\exists\,t>T:\,|\Q|>aT\big(\frac{t}{T}\big)^{\frac{1}{1+\gamma}}\bigg\}\bigg)=\\
&=1-\PP^{\ps[0]}\left(\left\{\Qo\in\real^3\mid|\Q[T]|\leq aT\right\}\right)+\\
&\phantom{\leq\PP^{\ps[0]}(\{\Qo\in\real^3}+\PP^{\ps[0]}\bigg(\bigg\{\Qo\in\real^3\mid
        |\Q[T]|\leq aT\quad\wedge\quad\exists\,t>T:\,|\Q|>aT\big(\frac{t}{T}\big)^{\frac{1}{1+\gamma}}\bigg\}\bigg)<\\
&<1-\left(\|\pspp[0]\|^2-\eps\right)\,.
\end{split}
\end{gather*}
Taken together this gives \eqref{eq.P(bound-in-slow-ball)-mixed} (again with $R:=aT$):
\begin{equation}
\label{eq.P(qt-slow)}
\begin{split}
\|\pspp[0]\|^2-\eps&<1-\PP^{\ps[0]}\left(\left\{\Qo\in\real^3\mid\exists\,t\geq T:
    \,|\Q|>aT\left(\frac{t}{T}\right)^{\frac{1}{1+\gamma}}\right\}\right)=\\
&=\PP^{\ps[0]}\left(\left\{\Qo\in\real^3\mid|\Q|\leq aT\left(\frac{t}{T}\right)^{\frac{1}{1+\gamma}}
        \quad\forall\,t\geq T\right\}\right)<\|\pspp[0]\|^2+\eps\,.
\end{split}
\end{equation}
\hfill\\

\noindent So to prove \eqref{eq.P(bound-in-slow-ball)-mixed} it is left to show \eqref{eq.P(Q<aT)-mixed}
and \eqref{eq.P(Q<aT-and-Q>)-mixed}. We start with \eqref{eq.P(Q<aT)-mixed}.
Note that for $a$ small and $T$ big enough  \eqref{eq.P(Q-bigger-aT)<-mixed} already gives us
\begin{multline*}
\PP^{\ps[0]}\left(\left\{\Qo\in\real^3\mid|\Q[T]|\leq aT\right\}\right)=
    1-\PP^{\ps[0]}\left(\left\{\Qo\in\real^3\mid|\Q[T]|>aT\right\}\right)>\\
>1-\|\psac[0]\|^2-\nhalf[\eps]=\|\pspp[0]\|^2-\nhalf[\eps]\,.
\end{multline*}
Similarly,
\begin{equation*}
\PP^{\ps[0]}\left(\left\{\Qo\in\real^3\mid|\Q[T]|\leq aT\right\}\right)<\|\pspp[0]\|^2+\nhalf[\eps]
\end{equation*}
for $a$ small and $T$ big enough follows directly from \eqref{eq.P(qt-in-B)>-mixed} and the fact that
\begin{equation*}
\left\{q\in\real^3\mid|\Qp[T]|\leq aT\right\}\subset\left\{q\in\real^3\mid\qtp[T]\not\in\settwoa\right\}=\G^{\rm c}
\end{equation*}
for all $\delta_1>0\,,\,\delta_2>0$ and $b>a$. Taken together this yields \eqref{eq.P(Q<aT)-mixed}:
\begin{equation*}
\|\pspp[0]\|^2+\nhalf[\eps]>\PP^{\ps[0]}\left(\left\{\Qo\in\real^3\mid|\Q[T]|\leq aT\right\}\right)
        >\|\pspp[0]\|^2-\nhalf[\eps]\,.
\end{equation*}
\\
Next we show \eqref{eq.P(Q<aT-and-Q>)-mixed}. Let $0<\gamma<2\alpha$. For convenience we define
\begin{equation*}
\A(a,T):=\bigg\{q\in\real^3\mid|\Qp[T]|\leq aT\:\wedge\:\exists t\geq T:\:
|\Qp|>aT\big(\frac{t}{T}\big)^{\frac{1}{1+\gamma}}\bigg\}\,.
\end{equation*}
Since $\Qp[t]$ (as a solution of the first order ODE \eqref{eq.eq-of-motion}) is continuous in $t$ $q\in\A(a,T)$
implies that $\Qp[t]$ crosses
the moving sphere $S_{\R}$ (with $\R:=(aT)\big(\frac{t}{T}\big)^{\frac{1}{1+\gamma}}$) at least once and
outwards in $[T,\infty)$. Therefore $\PP^{\ps[0]}\big(\A(a,T)\big)$ is bounded from above by the probability that some
trajectory crosses $S_{\R}$ in any direction in $[T,\infty)$. In Subsection 2.3.2 of \cite{berndl:94}
Berndl invoked the probabilistic meaning of the quantum probability current density
$J^\Psi=\left(j^{\ps[]}\,,\,|\ps|^2\right):=\left(\Im\big(\ps^*\nabla\ps\big)\,,\,|\ps|^2\right)$
to prove that the expected number of crossings of a smooth surface $\Sigma$ in configuration-space-time by
the random configuration-space-time trajectory $\left(\ro{Q}(\cdot,t)\,,\,t\right)$ is given by the flux across
this surface,
\begin{equation*}
\int\limits_{\Sigma}\left|J^\Psi(q,t)\cdot U\right|\,d\sigma\,,
\end{equation*}
where $U$ denotes the local unit normal vector at $(q,t)$\footnote{
This also includes tangential "crossings" in which the trajectory remains on the same side of $\Sigma$.}.
(See also the argument given in \cite{berndl:95}, p. 11.)
Since any trajectory $\left(\Qp\,,\,t\right)$ will cross $\Sigma$ an integral number of times (including $0$
and $\infty$) this expected value gives us an upper bound for the probability that $\left(\Qp\,,\,t\right)$
crosses $\Sigma$. In our case we obtain
\begin{equation*}
\PP^{\ps[0]}\left(\A(a,T)\right)\leq\int\limits_{\Sigma}\left|J^{\ps[]}(q,t)\cdot U\right|\,d\sigma=:\Pgamma(a,T)
\end{equation*}
where
\begin{equation*}
\Sigma=\left\{\left(\R\cos\phi\sin\theta,\R\sin\phi\sin\theta,\R\cos\theta,t\right)\mid (\phi,\theta,t)\in
    [0,2\pi)\times[0,\pi)\times[T,\infty)\right\}\,.
\end{equation*}
Then $U=\frac{1}{\sqrt{1+\partial_t\R}}\left(\hat{e}_r,\partial_t\R\right)$ and
$d\sigma=\sqrt{1+\partial_t\R}\R^2d\Omega\,dt$ where\\
$\hat{e}_r=\left(\cos\phi\sin\theta,\sin\phi\sin\theta,\cos\theta\right)$
is the usual unit normal vector of a three dimensional sphere and $d\Omega=\sin\theta\,d\phi\,d\theta$. We obtain
\begin{equation*}
\Pgamma(a,T)=\int\limits_{T}^{\infty}\,dt\int\limits_{S_{\R}}\left|j^\Psi(q,t)\cdot\hat{e}_r-
    |\ps(q)|^2\partial_t\R\right|\R^2\,d\Omega\,.
\end{equation*}

\noindent To control $\Pgamma(a,T)$ we need some estimates on $J^{\ps[]}$.

\begin{lem}
\label{lem.est-j}
Let $H=H_0+V$ with $V\in(V)_4$ and let zero be neither an eigenvalue nor a resonance of $H$. Let
$\ps[0]=\psac[0]+\pspp[0]$ with $\psac[0]\in\C$ and $\pspp[0]\in\D$ (for some $\alpha>0$).
Split the the quantum probability current density $J^{\ps[]}$ according to the splitting
of the wave function:
\begin{gather*}
\begin{split}
J^{\ps[]}(q,t)&=
    \left(j^{\ps[]}(q,t)\,,\,|\ps(q)|^2\right):=\left(\Im\big(\ps(q)^*\nabla\ps(q)\big)\,,\,|\ps(q)|^2\right)=\\
&=\left(j^{\pspp[]}(q,t)\,,\,|\pspp(q)|^2\right)+\left(j^{\psac[]}(q,t)\,,\,|\psac(q)|^2\right)
    +\left(j^{\rm m}(q,t)\,,\,\M(q,t)\right)
\end{split}\\
\intertext{with}
j^{\Psi^{\rm ac/pp}}(q,t):=\Im\big(\Psi_t^{\rm ac/pp}(q)^*\nabla\Psi_t^{\rm ac/pp}(q)\big)\,,\\
j^{\rm m}(q,t):=\Im\big(\pspp(q)^*\nabla\psac(q)+\psac(q)^*\nabla\pspp(q)\big)\\
\intertext{and}
\M(q,t):=2\Re\left(\pspp(q)^*\psac(q)\right)\,.
\end{gather*}
Then there is some $R>0$ such that for all $T>0$ there exists some $C<\infty$ such that
\begin{gather}
\label{eq.est-jpp}
\sup_{t\in\real}\big|j^{\pspp[]}(q,t)\big|\leq C|q|^{-3-2\alpha}\,,\quad
    \sup_{t\in\real}|\pspp(q)|^2\leq C|q|^{-3-2\alpha}\\
\intertext{for all $|q|>R$ and}
\label{eq.est-jac}
\big|j^{\psac[]}(q,t)\big|\leq C \big(t^{-3}+t^{-\nhalf}|q|^{-2}\big)\,,\quad
|\psac(q)|^2\leq C \big(t^{-3}+t^{-\nhalf}|q|^{-2}\big)\,,\\
\label{eq.est-jm}
\big|j^{\rm m}(q,t)\big|\leq C|q|^{-\nhalf-\alpha}\big(t^{-\nhalf}+t^{-1}|q|^{-1}\big)\,,\quad
\M(q,t)\leq C|q|^{-\nhalf-\alpha}\big(t^{-\nhalf}+t^{-1}|q|^{-1}\big)
\end{gather}
for all $|q|>R$ and $t\geq T$.
\end{lem}
\hfill\\
\noindent The proof of Lemma \ref{lem.est-j} can be found in the appendix.\\

\noindent With the help of Lemma \ref{lem.est-j} we shall show that $\Pgamma(a,T)<\nhalf[\eps]$ for $a$ small and
$T$ big enough.
We split
$\Pgamma(a,T)$ according to the splitting of $J^{\ps[]}$ in Lemma \ref{lem.est-j},
\begin{equation*}
\begin{split}
&\Pgamma(a,T)=
\Pgamma[pp](a,T)+\Pgamma[ac](a,T)+\Pgamma[m](a,T)\\
\intertext{with}
&\Pgamma[ac/pp](a,T)\negthickspace=\negthickspace
\int\limits_{T}^{\infty}dt\int\limits_{S_{\R}}\left|j^{\Psi^{\rm ac/pp}}(q,t)\cdot\hat{e}_r-
    |\Psi^{\rm ac/pp}(q)|^2\partial_t\R\right|\R^2\,d\Omega\\
\intertext{and}
&\Pgamma[m](a,T)\negthickspace=\negthickspace\int\limits_{T}^{\infty}dt\int
\limits_{S_{\R}}\left|j^{\rm m}(q,t)\cdot\hat{e}_r-\M(q,t)\partial_t\R\right|\R^2\,d\Omega\,,
\end{split}
\end{equation*}
and show that for $a$ small and $T$ big enough
\begin{gather*}
\Pgamma[pp](a,T)\leq C_1^aT^{-2\alpha},\quad\Pgamma[ac](a,T)\leq \tilde{C}a^2+C_2^aT^{-\nhalf[1]}\quad
\mbox{and}\quad\Pgamma[m](a,T)\leq C_3^aT^{-\alpha}
\end{gather*}
for some $\tilde{C}<\infty$ and some ($a$-dependent) $C_i^a<\infty$ ($i=1,\,2,\,3$).\\
By \eqref{eq.est-jpp} there is some $C<\infty$ such that
\begin{align*}
\Pgamma[pp](a,T)&\leq\int\limits_{T}^{\infty}\,dt\int\limits_{S_{\R}}\left[\left|j^{\pspp[]}(q,t)\right|+
    |\pspp(q)|^2\frac{1}{1+\gamma}\frac{\R}{t}\right]\R^2\,d\Omega\leq\\
&\leq 4\pi C\int\limits_{T}^{\infty}\left[\R^{-1-2\alpha}+\frac{\R^{-2\alpha}t^{-1}}{1+\gamma}\right]\,dt=
    4\pi C\left[\frac{1+\gamma}{2\alpha-\gamma}a^{-1-2\alpha}+\frac{a^{-2\alpha}}{2\alpha}\right]T^{-2\alpha}=\\
&= C_1^aT^{-2\alpha}\,.
\end{align*}
In exactly the same way we get the desired bounds on $\Pgamma[ac](a,T)$ and $\Pgamma[m](a,T)$ since, for $T$ and
$|q|=\R\geq R_{a,T}(T)$
big enough, i.e. for $T$ big enough, \eqref{eq.est-jac} resp. \eqref{eq.est-jm} implies
\begin{gather*}
\begin{split}
\Pgamma[ac](a,T)\leq 4\pi C \int\limits_{T}^{\infty}\left[\R^2t^{-3}+t^{-\nhalf}+
    \frac{1}{1+\gamma}\left(\R^3t^{-4}+\R t^{-\nhalf[5]}\right)\right]\,dt\leq \tilde{C}a^2+C_2^aT^{-\nhalf[1]}
\end{split}
\intertext{resp.}
\begin{split}
&\Pgamma[m](a,T)\leq\\
&\leq4\pi C \int\limits_{T}^{\infty}\left[\R^{\nhalf[1]-\alpha}t^{-\nhalf}+\R^{-\nhalf[1]-\alpha}t^{-1}+
    \frac{1}{1+\gamma}\left(\R^{\nhalf-\alpha}t^{-\nhalf[5]}+\R^{\nhalf[1]-\alpha}t^{-2}\right)\right]\,dt=\\
&\phantom{\leq4}\leq C_3^aT^{-\alpha}\,.
\end{split}
\end{gather*}
Thus we have proved \eqref{eq.P(Q<aT-and-Q>)-mixed} and can proceed to show $(ii)$.\\

\noindent It is now easy to prove that $v_\infty(\Qo)$ exists for $\PP^{\ps[0]}$-almost all $\Qo\in\real^3$.
Note that $|\Qp|\leq aT\big(\frac{t}{T}\big)^{\frac{1}{1+\gamma}}$ for all $t\geq T$ implies
$v_\infty(q):=\lim\limits_{t\rightarrow\infty}\qtp=0$. So by \eqref{eq.settwoa-good-mixed} $v_\infty(\Qo)$ exists
for all initial configurations $\Qo$ in
\begin{equation*}
\left\{\Qo\in\real^3\mid|\Q|\leq aT\big(\frac{t}{T}\big)^{\frac{1}{1+\gamma}}\quad\forall t\geq T\right\}
\cup\left\{\Qo\in\real^3\mid\qt[T]\in\settwoa\right\}\,.
\end{equation*}
Since the latter two sets are disjoint and have measure arbitrary close to $\|\pspp[0]\|^2$ (by \eqref{eq.P(qt-slow)})
and $\|\psac[0]\|^2$ (by \eqref{eq.P(qt-in-B)>-mixed}) respectively we get almost sure existence of $v_\infty$.\\

\noindent Finally it is left to show \eqref{eq.v-infty-distr-mixed}. Let $A\subset\real^3$ be measurable. By dominated
convergence, equivariance \eqref{eq.equivariance} and \eqref{eq.est-psac-l2} we get (for details see the
proof of \eqref{eq.v-infty-distr-pure} of Theorem \ref{thm.pure-scattering})
\begin{equation}
\label{eq.P(v-infty-in-A)-mixed}
\begin{split}
\PP^{\ps[0]}\big(v_\infty\in A\big)&=
        \lim\limits_{t\rightarrow\infty}\PP^{\ps[0]}\big(\big\{q\in\real^3\mid\;\qtp\in A\big\}\big)=\\
&=\lim\limits_{t\rightarrow\infty}\left[\big\|\chi_{\xt\in A}\phi_1(\cdot,t)\big\|^2+
        \big\|\chi_{\xt\in A}\pspp\big\|^2+
        2\Re\left\langle\chi_{\xt\in A}\phi_1(\cdot,t)|\pspp\right\rangle\right].
\end{split}
\end{equation}
The first term yields
\begin{equation}
\label{eq.v-infty-distr-sc-part-mixed}
\big\|\chi_{\xt\in A}\phi_1(\cdot,t)\big\|^2=\int\limits_A|\psat(k)|\,d^3k\,.
\end{equation}
The third term tends to zero as $t\rightarrow\infty$: With $|\Re(z)|\leq|z|$ and Schwarz inequality we
get for every $\gamma>0$
\begin{align*}
\bigg|\Re\bigg\langle\chi_{\xt\in A}\phi_1(\cdot,t)&|\pspp\bigg\rangle\bigg|\leq\\
&\leq\left|\Re\left\langle\chi_{\xt\in A\wedge|q|\leq t^{\frac{1}{1+\gamma}}}\phi_1(\cdot,t)|\pspp\right\rangle+
        \Re\left\langle\phi_1(\cdot,t)|\chi_{\xt\in A\wedge|q|> t^{\frac{1}{1+\gamma}}}\pspp
                \right\rangle\right|\leq\\
&\leq\big\|\chi_{\xt\in A\wedge|q|\leq t^{\frac{1}{1+\gamma}}}\phi_1(\cdot,t)\big\|\,\big\|\pspp[0]\big\|+
        \big\|\phi_1(\cdot,t)\big\|\,\big\|\chi_{\xt\in A\wedge|q|> t^{\frac{1}{1+\gamma}}}\pspp\big\|\,.
\end{align*}
By \eqref{eq.v-infty-distr-sc-part-mixed}
\begin{gather*}
\big\|\phi_1(\cdot,t)\big\|=\big\|\psat\big\|=\big\|\psac[0]\big\|\\
\intertext{and}
\big\|\chi_{\xt\in A\wedge|q|\leq t^{\frac{1}{1+\gamma}}}\phi_1(\cdot,t)\big\|^2=
        \int\limits_{|k|\leq t^{\frac{-\gamma}{1+\gamma}}}|\psat(k)|\,d^3k\,.
\end{gather*}
This tends to zero as $t\rightarrow\infty$ since by Lemma \ref{lem.mapping} $\psat(k)$ is bounded outside $k=0$.
Moreover,
\begin{equation*}
\lim\limits_{t\rightarrow\infty}\big\|\chi_{\xt\in A\wedge|q|> t^{\frac{1}{1+\gamma}}}\pspp\big\|=0
\end{equation*}
by \eqref{eq.est-pspp-l2}.\\
Thus it is left to show that the second term in the last line of \eqref{eq.P(v-infty-in-A)-mixed} yields
\begin{equation}
\label{eq.chi-A-pspp}
\lim\limits_{t\rightarrow\infty}\big\|\chi_{\xt\in A}\pspp\big\|^2=
    \begin{cases}
    \lim\limits_{t\rightarrow\infty}\big\|\pspp\big\|^2=\big\|\pspp[0]\big\|^2& \text{if $0\in A$,}\\
    0& \text{if $0\not\in A$.}
    \end{cases}
\end{equation}
Again using \eqref{eq.est-pspp-l2} we get
\begin{equation*}
\lim\limits_{t\rightarrow\infty}\big\|\chi_{\xt\in A}\pspp\big\|^2=
    \lim\limits_{t\rightarrow\infty}\big\|\chi_{\xt\in A\wedge|q|\leq t^{\frac{1}{1+\gamma}}}\pspp\big\|^2.
\end{equation*}
Since $|q|\leq t^{\frac{1}{1+\gamma}}$ implies $\lim\limits_{t\rightarrow\infty}\frac{|q|}{t}=0$, we have
\begin{equation*}
\lim\limits_{t\rightarrow\infty}\chi_{\xt\in A\wedge|q|\leq t^{\frac{1}{1+\gamma}}}(q)=
    \begin{cases}
    \lim\limits_{t\rightarrow\infty}\chi_{|q|\leq t^{\frac{1}{1+\gamma}}}(q)=1& \text{if $0\in A$,}\\
    0& \text{if $0\not\in A$}
    \end{cases}
\end{equation*}
for all $q\in\real^3$. Moreover, $|\chi_{\xt\in A\wedge|q|\leq t^{\frac{1}{1+\gamma}}}\pspp(q)|^2
\leq|\pspp(\Qo)|^2$, so by dominated convergence we finally get \eqref{eq.chi-A-pspp}.
\end{pro}
\hfill\\

\begin{pro}[of Corollary \ref{cor.straight-mixed}]\\
The proof of Corollary \ref{cor.straight-mixed} is completely analogous to that of
Corollary \ref{cor.straight-pure}.
\end{pro}

%-------------------------------------------------------------------------------------------
\section{Appendix}
\label{sec.app}
\begin{pro}[of Lemma \ref{lem.est-psac-phi3}]
The estimates in \eqref{eq.est-phi3} and \eqref{eq.est-Delta-phi3} were done by Teufel, D\"urr and Berndl
in \cite{teufel2:99} (Equations (15) and (16)). Their $\beta$ is our $\phi_3$. Furthermore rather than $\ps[0]\in\C$
they used conditions on $\psat$ included in $\psat\in\hat{\C}$ to prove their Equations (15)
and (16). Because of Lemma \ref{lem.mapping} this poses no problems.
\end{pro}
\hfill\\

\begin{pro}[of Lemma \ref{lem.est-psac-phi1}]
Keeping in mind that $\psout$ evolves according to the free time evolution, i.e.
\begin{equation*}
\psout=(2\pi it)^{-\nhalf}\int\limits_{\real^3}e^{i\frac{|q-y|^2}{2t}}\psout[0](y)\,d^3y
=(2\pi)^{-\nhalf}\int\limits_{\real^3}e^{i(k\cdot q-\nhalf[k^2t])}\psat(k)\,d^3k\,,
\end{equation*}
\eqref{eq.psout-phi1-phi2} is a straightforward calculation.\\
The estimates in \eqref{eq.est-phi2} and \eqref{eq.est-Delta-phi2} were done by D\"urr, Moser and Pickl
in \cite{duerr1:04} (Equations (17) and (18)). Their $\alpha$ is our $\psout[]$.\\
To prove \eqref{eq.est-psac-l2} we use that by \eqref{eq.psac-psout-phi3} and \eqref{eq.psout-phi1-phi2}
\begin{equation*}
\|\psac-\phi_1(\cdot,t)\|=\|\psac-\psout\|+\|\phi_2(\cdot,t)\|.
\end{equation*}
For the first term we use the definition of the outgoing asymptote $\psout=W_+^{-1}\psac$ and get
\begin{align*}
\lim\limits_{t\rightarrow\infty}\|\psac-\psout\|&=
\lim\limits_{t\rightarrow\infty}\|e^{-iHt}\psac[0]-e^{-iH_0t}\psout[0]\|=\\
&=\lim\limits_{t\rightarrow\infty}\|\ps[0]-e^{iHt}e^{-iH_0t}\psout[0]\|=\|\psac[0]-W_+\psout[0]\|=0.
\end{align*}
The estimation of the second term is also standard (see e.g. \cite{dollard:69,duerr:01}).
\begin{equation*}
\begin{split}
\|\phi_2(\cdot,t)\|&= \|(2\pi t)^{-\nhalf}\int\limits_{\real^n}e^{-i\frac{\cdot}{t}\cdot y}
        (e^{i\frac{y^2}{2t}}-1)\psout[0](y)\,d^ny\|
        =\|\mathcal{F}\bigl((e^{i\frac{y^2}{2t}}-1)\psout[0](y)\big)(\cdot)\|=\\
    &=\|(e^{i\frac{\cdot^2}{2t}}-1)\psout[0]\|
\end{split}
\end{equation*}
and $(e^{i\frac{q^2}{2t}}-1)\psout[0](q)\rightarrow0\,$ pointwise as $t\rightarrow\infty$. Moreover,
$|(e^{i\frac{q^2}{2t}}-1)\psout[0](q)|^2\leq4|\psout(q)|^2\in L_1(\real^3)$, so
\begin{equation*}
\lim\limits_{t\rightarrow\infty}\|\phi_2(\cdot,t)\|=0
\end{equation*}
by dominated convergence.
\end{pro}
\hfill\\

\begin{pro}[of Lemma \ref{lem.est-j}]
Let $T>0$. Since $|j^{\pspp[]}(q,t)|=|\Im\big(\pspp(q)^*\nabla\pspp(q)\big)|\leq|\pspp(q)||\nabla\pspp(q)|$
\eqref{eq.est-jpp} immediately follows from Definition \ref{def.pspp-less-q^(-3)}.\\
To get bounds on
\begin{gather*}
j^{\psac[]}(q,t)=\Im\big(\psac(q)^*\nabla\psac(q)\big)\quad\mbox{resp.}\quad |\psac(q)|^2\\
\intertext{and}
j^m(q,t)=\Im\big(\pspp(q)^*\nabla\psac(q)+\psac(q)^*\nabla\pspp(q)\big)\quad\mbox{resp.}\quad
    \M(q,t)=2\Re\left(\pspp(q)^*\psac(q)\right)
\end{gather*}
we need to estimate $\psac[]$ and $\nabla\psac[]$. According to Lemma \ref{lem.mapping} there exits some
$C<\infty$ such that $|\psat(k)|\leq C|k|^{-r}$ for all $r\in\{0,1,\ldots,5\}$. Therefor,
using repeatedly Definition \ref{def.pspp-less-q^(-3)}, Lemma \ref{lem.est-psac-phi3} and Lemma \ref{lem.est-psac-phi1}
we obtain for all $t\geq T$, $|q|$ big enough, $r\in\{0,1,\ldots,5\}$
and some suitable $C_T<\infty$
\begin{gather*}
\begin{split}
|\psac(q)|&\leq|\phi_1(q,t)|+|\phi_2(q,t)|+|\phi_3(q,t)|\leq
        t^{-\nhalf}\left|\psat\left(\xt\right)\right|+Ct^{-2}+\frac{C_T}{|q|(t+|q|)}\leq\\
    &\leq C_T\left(t^{-\nhalf}\left(\tx\right)^r+t^{-2}+\frac{1}{|q|(t+|q|)}\right)
\end{split}\\
\intertext{and}
\begin{split}
|\nabla\psac(q)|\leq\left|\nabla\psout(q)-i\xt\phi_1(q,t)\right|+\frac{|q|}{t}&|\phi_1(q,t)|+|\phi_3(q,t)|\leq\\
     &\leq C_T\left(t^{-\nhalf}\left(\tx\right)^{r-1}+t^{-2}+\frac{1}{|q|(t+|q|)}\right).
\end{split}
\end{gather*}
For $r=0$, resp. $r=1$ this and Definition \ref{def.pspp-less-q^(-3)} yield \eqref{eq.est-jac} and \eqref{eq.est-jm}.
\end{pro}

%References---------------------------------------------------------
\bibliographystyle{/home/math/roemer/biblist/literatur}
\bibliography{/home/math/roemer/biblist/literatur}
\end{document}